\documentclass[journal,english,twoside]{IEEEtran}

\usepackage{balance}

\usepackage{cite}

\usepackage[pdftex]{graphicx}

\usepackage[caption=false, font=footnotesize]{subfig}

\usepackage[cmex10]{amsmath}
\usepackage{amssymb}
\usepackage{mathrsfs}
\usepackage{bigints}

\usepackage{siunitx}

\usepackage{multicol}
\usepackage{lipsum}

\usepackage{stfloats}

\newcounter{storeeqcounter}
\newcounter{tempeqcounter}

\usepackage{url}

\usepackage{todonotes}
\definecolor{morange}{rgb}{0.8,0.2,0}
\definecolor{mblue}{rgb}{0,0,0}
\definecolor{mgreen}{rgb}{0,0,0}
\definecolor{mred}{rgb}{1,0,0}

\newcommand{\tbirkan}[1]{{\color{mblue}{#1}}}
\newcommand{\tbayram}[1]{{\color{mgreen}{#1}}}

\DeclareMathOperator{\erfc}{erfc}

\newcommand{\fhit}[2]{F_{\text{hit}}(t, #1, #2)}
\newcommand{\fhitx}[2]{F_{\text{hit}}(t_s, #1, #2)}
\newcommand{\pointA}{A}
\newcommand{\pointApr}{\pointA^{\prime}}
\newcommand{\fhitA}[1]{F_{\text{hit}}^{#1}(t)}
\newcommand{\pthetaT}[1]{p(\theta,#1)}
\newcommand{\ptheta}{p(\theta)}
\newcommand{\ntheta}{N_{\theta}}
\newcommand{\epstheta}{\epsilon \!\left( \theta  \right)}

\newcommand{\afactor}{T(\theta)}
\newcommand{\tfactor}{\phi(t)}
\newcommand{\tfactort}[1]{\phi(#1)}

\newcommand{\optangle}{\alpha^{*}}

\newcommand{\tsym}{t_s}
\newcommand{\ntxAmp}[1]{N_{#1}^{\text{Tx}}}
\newcommand{\symi}[1]{s_{#1}}

\newcommand{\stxK}[1]{s[#1]}
\newcommand{\shatK}[1]{\hat{s}[#1]}

\newcommand{\merfc}[1]{\mbox{erfc}\!\left(\! #1 \! \right)\!}

\newcommand{\xth}[1]{{#1}^{th}}

\title{Molecular Signal Modeling of a Partially Counting Absorbing Spherical Receiver}

\author{Bayram Cevdet Akdeniz,~\IEEEmembership{Student Member,~IEEE}, Nafi Ahmet Turgut,~\IEEEmembership{Student Member,~IEEE}, \\ 
H. Birkan Yilmaz,~\IEEEmembership{Member,~IEEE}, Chan-Byoung Chae,~\IEEEmembership{Senior Member,~IEEE},\\ Tuna Tugcu,~\IEEEmembership{Member,~IEEE}, and Ali Emre Pusane,~\IEEEmembership{Member,~IEEE} 

\thanks{B.C. Akdeniz,  and A. E. Pusane are with the Department of Electrical and Electronics Engineering, Bogazici University, Istanbul, 34342, Turkey (e-mail: bayram.akdeniz@boun.edu.tr and ali.pusane@boun.edu.tr).}
\thanks{ N.A. Turgut was with the Department of Electrical and Electronics Engineering, Bogazici University. He is now with the Electronics Engineering in Koc Univesity  (e-mail: nturgut17@ku.edu.tr).}
\thanks{T. Tugcu is with NETLAB, Department of Computer Engineering, Bogazici University, Istanbul, 34342, Turkey (e-mail: tugcu@boun.edu.tr).}
\thanks{H. B. Yilmaz was with the Yonsei Institute of Convergence Technology, School of Integrated Technology, Yonsei University, Korea. He is now with the Department of Telematics Engineering, Universitat Politecnica de Catalunya, Barcelona, Spain (e-mail: birkan.yilmaz@upc.edu).}
\thanks{C.-B. Chae is with the Yonsei Institute of Convergence Technology, School of Integrated Technology, Yonsei University, Korea (e-mail:
cbchae@yonsei.ac.kr).}
}

\begin{document}
\maketitle

\begin{abstract}
\tbirkan{To communicate at the nanoscale, researchers have proposed molecular communication as an energy-efficient solution. The drawback to this solution is that the histogram of the molecules\textsc{'} hitting times, which constitute the molecular signal at the receiver, has a heavy tail. Reducing the effects of this heavy tail, inter-symbol interference (ISI), has been the focus of most prior research. In this paper, a novel way of decreasing the ISI by defining a counting region on the spherical receiver's surface facing towards the transmitter node is proposed. The beneficial effect comes from the fact that the molecules received from the back lobe of the receiver are more likely to be coming through longer paths that contribute to ISI. In order to justify this idea, the joint distribution of the arrival molecules with respect to  angle and time is derived. Using this distribution, the channel model function is approximated for the proposed system, i.e., the partially counting absorbing spherical receiver. After validating the channel model function, the characteristics of the molecular signal are investigated and improved performance is presented. Moreover, the optimal counting region in terms of bit error rate is found analytically.}
\end{abstract}


\section{Introduction}
\IEEEPARstart{T}{hrough} billions of years of producing communication at small scales (i.e., distances of up to a few micro/nano meters), nature has provided, tested, and improved molecular communication (MC). Humans, on the other hand, struggle at this scale to utilize electromagnetic waves thanks to the constraints imposed by the ratio of the antenna size to the wavelength of the electromagnetic signal~\cite{akyildiz2008nanonetworksAN,farsad2016comprehensiveSO}. As an alternative to electromagnetic signal, molecular signals have been proposed for nanonetworks in order to eliminate antenna constraint. There are other advantages to molecular communication-- molecular signals are typically more bio-compatible and can reach an intended receiver within challenging environments even on the macro-scale such as pipelines, tunnels, and saline water environments~\cite{guo2015molecularVE}. Therefore, researchers direct their attentions to molecular communication to achieve communication in nanonetworks.

Most of the existing research on MC has focused on channel modeling, interference mitigation, and modulation issues ~\cite{genc2016isiAM,kuran2011modulationTF,noel2014improvingRP,Guo_WC16}. To address the challenges in a methodological and inclusive manner, IEEE has established the standardization group IEEE P1906.1 for MC~\cite{IEEEP1906_1}. 

One of the main challenges in MC is to develop valid channel models capable of representing a time-dependent received signal. For the receiver and the reception process in diffusion-based MC models,  there are mainly two types of models-- the passive and absorbing receivers. The former assumes the molecules are unaffected by the receiver while the latter assumes the molecules are absorbed whenever they hit the receiver. In the passive receiver case, the molecules can pass through the receiver node surface multiple times without interaction~\cite{pierobon2010physicalET,kilinc2013receiverDF,noel2014improvingRP}. Therefore, the molecules are allowed to contribute to the received signal multiple times when the receiver is passive. For the absorbing receiver case, the molecules contribute to the received signal only once and the molecules that hit the receiver are removed from the environment~\cite{srinivas2012molecularCI_inverseG,nakano2012channelMA_oneD_COML,yilmaz2014threeDC,Yilmaz_WCL_17}. This process is modeled by the first-passage process and to model the received signal, we focus on the time-dependent first hitting histogram~\cite{redner2001guide}. In~\cite{srinivas2012molecularCI_inverseG} and \cite{nakano2012channelMA_oneD_COML}, the received molecular signal is modeled in a one-dimensional (1-D) environment with an absorbing receiver and the system performance is analyzed by utilizing the received signal model at the physical layer. In~\cite{yilmaz2014threeDC}, a received signal model is introduced for a point transmitter and a spherical absorbing receiver in a 3-D environment. Since then, researchers have focused on modeling the received signal for an absorbing receiver while relaxing some of the assumptions. Instead of using a fully absorbing receiver, the authors have incorporated the receptor effect instead of using a fully absorbing receiver~\cite{akkaya2015effectOR_receptor_COML}. Similarly, researchers have utilized machine learning techniques to model the received signal for a spherical reflecting transmitter with single absorbing receiver~\cite{yilmaz2016machineLA} or multiple point transmitters with multiple absorbing receivers~\cite{lee2017machineLB_ARXIV, moleMIMO}. {In \cite{noel2016channel}, the communication between a spherical receiver and a spherical transmitter in which the surface is covered with evenly-spaced point transmitters has been modeled and the channel impulse response has been presented.}

{In addition to channel modeling, another major and common challenge in MC systems is the inter-symbol interference (ISI), which is caused due to the late reception of some messenger molecules in the channel. Consequently, many recent works have focused on overcoming this issue by proposing either modulation or equalization methods. In \cite{kabir2015d} and \cite{arjmandi2013diffusion}, the authors have used, simultaneously, different types of molecules as two orthogonal channels. Trying to cope with ISI molecules, the authors in \cite{tepekule2015novel} have released an additional type of messenger molecule. To reduce the ISI effect, the receiver observes and evaluates the difference in the number of molecules of both types. Although these works show promising results for reducing ISI, they require the usage of different types of molecules, hence different types of receptors at the receiver causing increased complexity of the system. 

There are also other solutions that use one type of messenger molecule. For instance in \cite{tepekule2015isi},  ISI is used as a constructive component by adjusting the  number of released molecules so that the residual molecules lead to a beneficial effect on decoding of the following symbols. In \cite{kilinc2013receiver}, conventional equalization methods like minimum mean square error (MMSE), decision feedback equalizers, and maximum likelihood sequence estimation methods are proposed for MC channels. While these methods have incremental effects on the channel, they require a significant amount of additional computational complexity.}

In this paper, we model the received molecular signal for a partially counting absorbing receiver. That is, the receiver absorbs all hitting molecules but those counted are only the ones hitting at a specific site. Modeling the time-dependent received signal for such a system is an open issue and has the potential to enhance the communication system performance without {any significant additional cost}. Most of the received molecules are absorbed from the surface area facing towards the transmitter side. As the path to the back side of the receiver is longer, the receptors on that side are more likely to receive the interference molecules. Therefore, limiting the counting area to the front side with a limited surface area enhances the signal quality. The main contributions of this paper are listed as follows:
\begin{itemize}

\item The derivation of the joint distribution of the received molecules with respect to time and angle 
\item The modelling the received signal
\item The investigating of the signal properties, and
\item The finding of the optimal region for counting
\end{itemize}
for a partially counting and absorbing spherical receiver in a diffusion-based MC system.


\section{System Model\label{sec_system_model}}
The received signal in a diffusion-based MC system is affected by  three main processes: emission, propagation, and reception. Analytical derivations for the channel model depend on the emitter, the receiver, the environment, and the propagation dynamics. Therefore, we give the details of the system before deriving the channel model.

\subsection{Topology Model}
\begin{figure}[!t]
	\begin{center}
		\includegraphics[width=0.99\columnwidth,keepaspectratio]%
		{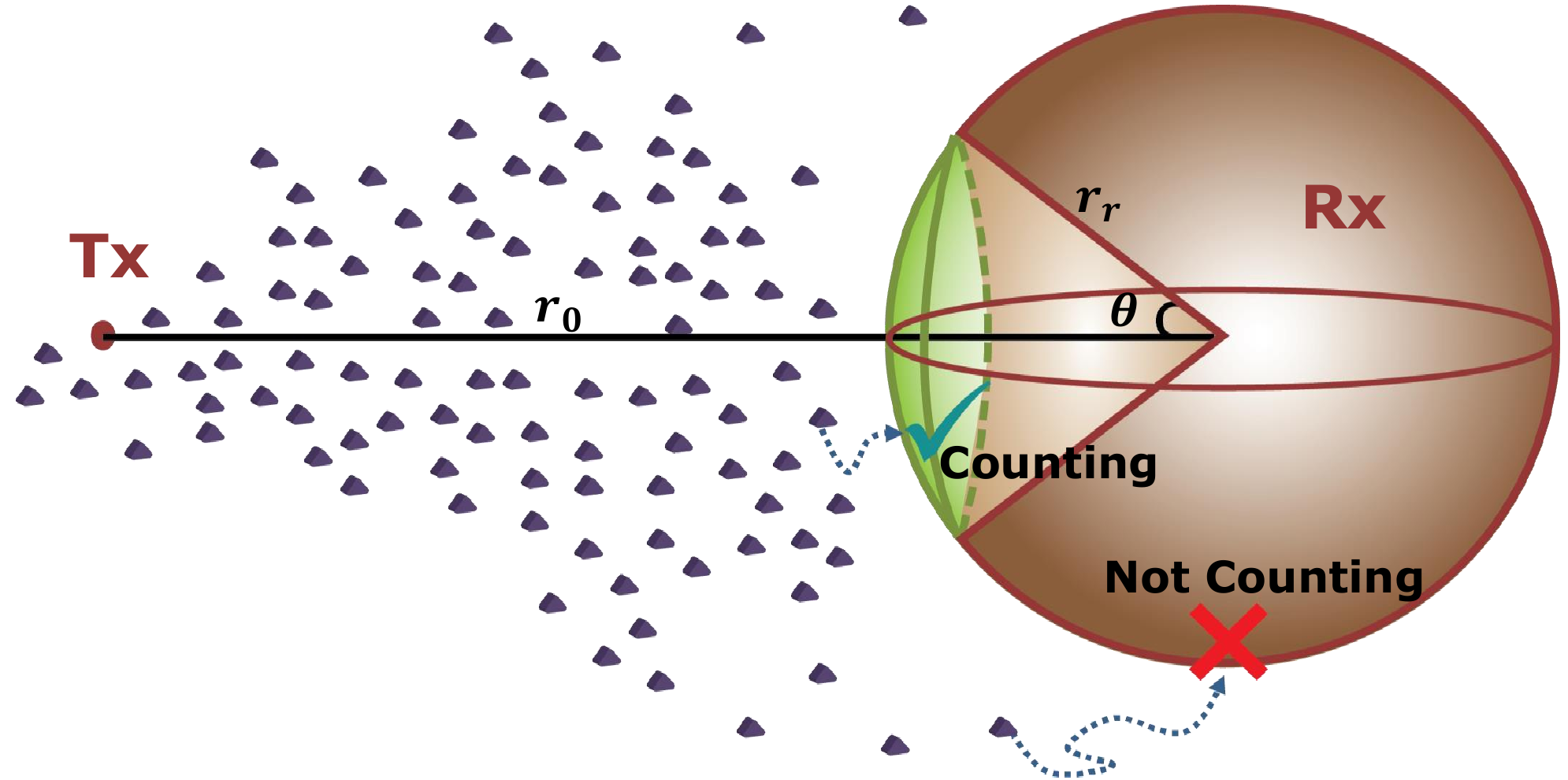}
		\caption{System model of a diffusion-based MC with a point transmitter and a partially counting absorbing spherical receiver.}
		\label{fig_system_model}
	\end{center}
\end{figure}

We consider a diffusion-based MC system with one point transmitter and one spherical receiver in a 3-D environment (Fig.~\ref{fig_system_model}). Novel feature of the receiver (Rx) is the ability to count the molecules absorbed only through a specific region. This feature complicates the modeling procedure of the received signal. In Fig.~\ref{fig_system_model}, the circular cap facing towards the transmitter node (Tx) counts the absorbed molecules while the rest of the surface area absorbs but does not count the molecules. 

As shown in Fig.~\ref{fig_system_model}, the molecules propagate by the diffusion process when they are emitted from the Tx point. The distance between the emission point and the center of the receiver is denoted by $r_0$ and the radius of the absorbing receiver is denoted by $r_r$. The circular cap that counts is determined by the $\theta$ angle, which we name as the \emph{counting region}.

It is assumed that Tx and Rx nodes are fully synchronized in the time domain, and the interactions between diffusing molecules are ignored. No environmental or counting circuit noise is considered; only the diffusion noise is considered to isolate the signaling gain due to partial counting receiver. Furthermore, it is assumed that a mechanism in Rx node determines the direction of Tx and aligns its \emph{counting region} facing towards Tx.

\subsection{Diffusion Model}
The emitted molecules propagate subject to Brownian Motion, which is described by the Wiener process~\cite{srinivas2012molecularCI_inverseG}. 
The Wiener process $W(t)$ is characterized as follows:
\begin{itemize}
\item $W(0) = 0$,
\item $W(t)$ is almost surely continuous, 
\item $W(t)$ has independent increments,
\item $W(t_2) - W(t_1) \sim \mathscr{N}(0,\,c(t_2-t_1))$ for $0\leq t_1 \leq t_2$
\end{itemize}
is the Gaussian distribution with mean $\mu$ and variance $\sigma^2$ and $c$ is a constant. Simulating the Brownian Motion includes consecutive steps in an $n$-dimensional space that obeys Wiener process dynamics. For an accurate simulation, time is divided into sufficiently small time intervals ($\Delta t$), and at each time interval the molecules take random steps in all dimensions.
In an $n$-dimensional space, a random step is given as
\begin{align}
\begin{split}
\Delta \zeta &= (\Delta \zeta_1, ..., \Delta \zeta_n), \\
\Delta \zeta_i &\sim \mathscr{N}(0, \, 2D\Delta t) \;\; \forall i \in \{1,..,n\},
\end{split}
\label{brown}
\end{align}
where $\Delta \zeta$, $\Delta \zeta_i$, and $D$ correspond to the random displacement vector, the displacement at the $\xth{i}$ dimension, and the diffusion coefficient, respectively.

\subsection{Modulation \& Demodulation}
In this paper, concentration shift keying (CSK) based modulation technique is used. General form of CSK is introduced in~\cite{kuran2011modulationTF,nrkim_mod}. In CSK based modulation techniques, the information is modulated on the amount of the transmitted molecules at the start of each symbol duration ($\tsym$).
\begin{figure}[!t]
	\begin{center}		\includegraphics[width=0.99\columnwidth,keepaspectratio]%
		{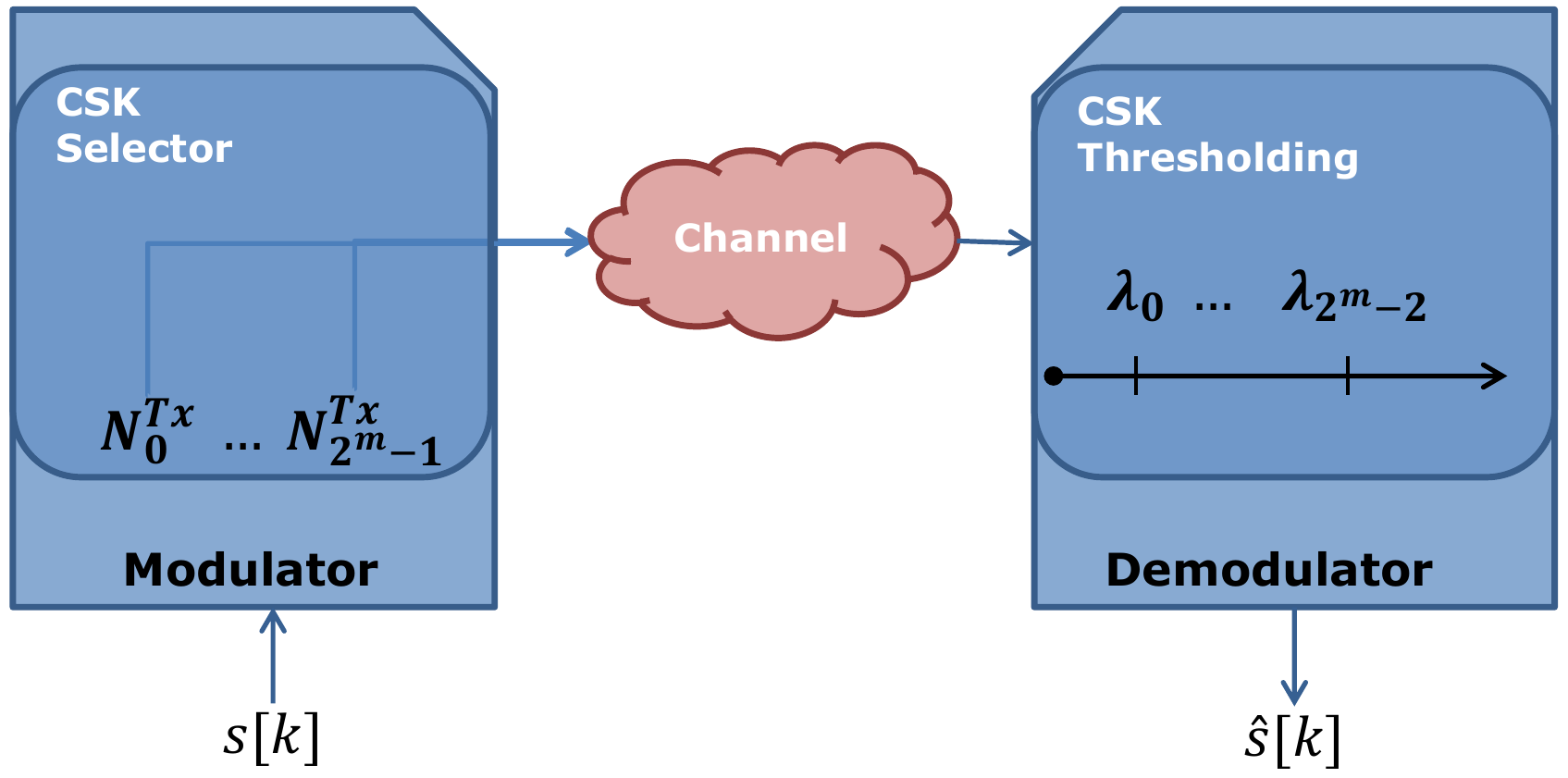}
		\caption{Concentration-based modulator and demodulator.}
		\label{fig_modulation}
	\end{center}
\end{figure}

General structure of CSK based modulations is depicted in Fig.~\ref{fig_modulation}. For the $\xth{k}$ symbol $\stxK{k}$, the modulator maps the symbol to the amount of molecules to emit (i.e., $\xth{i}$ symbol is mapped to $\ntxAmp{i}$) at the start of the $\xth{k}$ symbol duration. Depending on the modulation order ($m$), the number of possible symbols is determined and equals to $2^m$. In this paper, we use binary CSK where $m\!=\!1$, i.e., it has two symbols $\symi{0}$ and $\symi{1}$ which represent bit-0 and bit-1, respectively. After CSK selector maps the symbol to the amount, the molecules are emitted to the channel and they propagate by diffusion. During the symbol duration, the arriving molecules are absorbed and counted according to the counting logic. At the end of the $\xth{k}$ symbol duration, the final value is thresholded for obtaining the detected symbol $\shatK{k}$ for the $\xth{k}$ symbol.

\section{Channel Model for Partial Counting Receiver\label{sec_channel_model}}

The joint cumulative angle and time distribution of released molecules by a point transmitter at the spherical receiver deserves an analytical derivation. This distribution function is utilized to determine the partial channel taps analytically.

In the literature, marginal cumulative distribution with respect to time is derived for a fully absorbing spherical receiver and introduced to the MC domain from a communication perspective~\cite{yilmaz2014threeDC} as
\begin{equation}
\fhit{r_0}{r_r} =\cfrac { { r }_{ r } }{ { r }_{ 0 } } \, \merfc{ \cfrac { { r }_{ 0 }-{ r }_{ r } }{ \sqrt { 4Dt }  }  }, 
\label{fhit}
\end{equation}
where $\merfc{ \cdot }$ is the  complementary error function. Furthermore, the marginal angular distribution of the molecules is given in \cite{redner2001guide} (6.3.3a) for a ${\mbox{3-D}}$ medium when the time goes to infinity as 
\begin{equation}
\ptheta = 2\pi { r }_{ r }^{ 2 }\sin { \theta }  \,\, \epstheta,
\label{eq:teta}
\end{equation}
where 
\begin{align}
\epstheta = 
	\frac{ \left( 1-\cfrac { { r }_{ r }^{ 2 } }{ { r }_{ 0 }^{ 2 } }  \right)  }
    { 4\pi { r }_{ r }{ r }_{ 0 }{ \left( 1-\cfrac { 2{ r }_{ r } }{ { r }_{ 0 } } \cos { \theta  } +\cfrac { { r }_{ r }^{ 2 } }{ { r }_{ 0 }^{ 2 } }  \right)  }^{ { 3 }/{ 2 } }}.
\end{align}

\begin{figure}[!t]
	\begin{center}
		\includegraphics[width=0.95\columnwidth,keepaspectratio]%
		{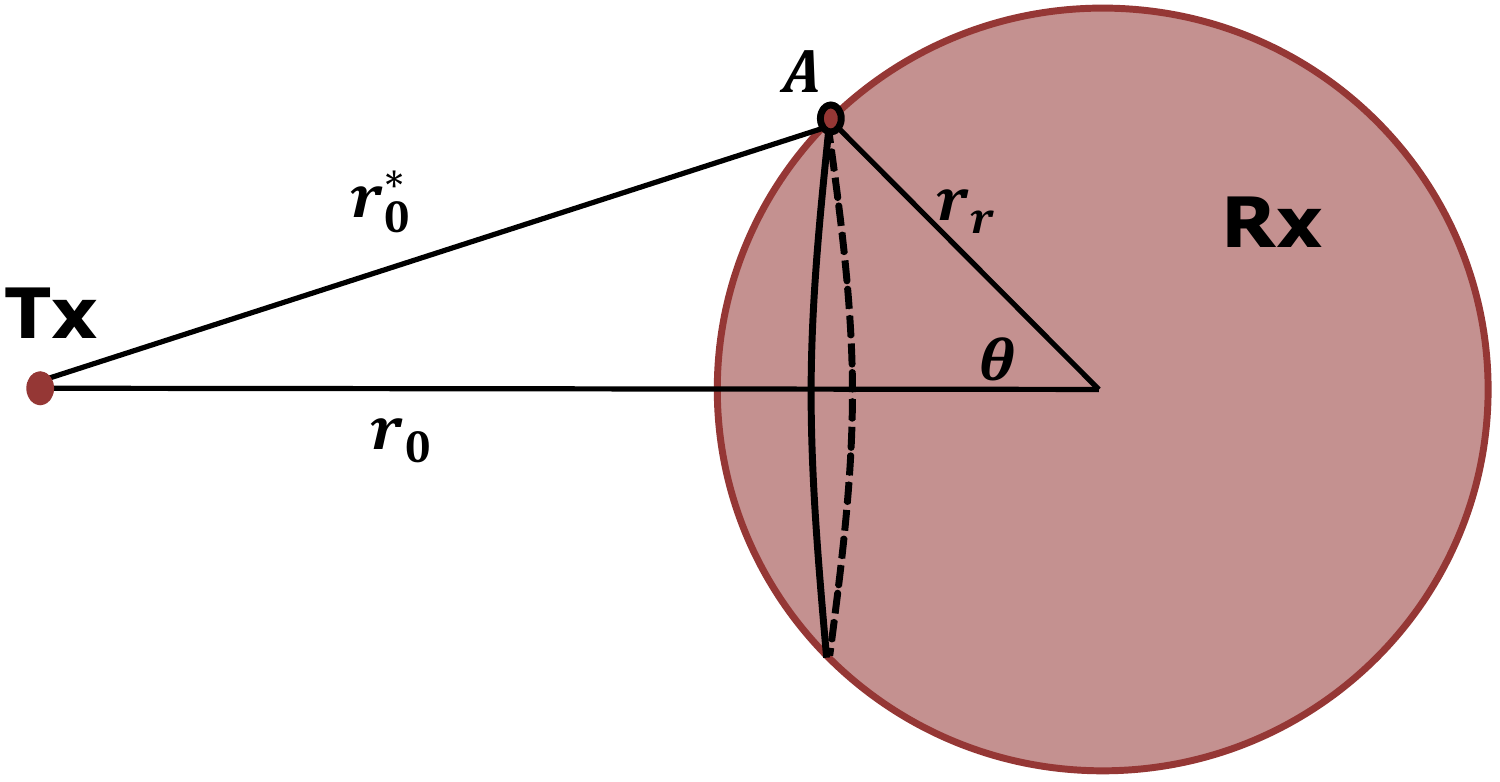}
		\caption{An infinitesimally small sphere over the circular region on the surface of the sphere. The circular region is determined by the angle $\theta$.}
		\label{fig_fig1}
	\end{center}
\end{figure}

\begin{figure}[!t]
	\begin{center}
		\includegraphics[width=0.99\columnwidth,keepaspectratio]%
		{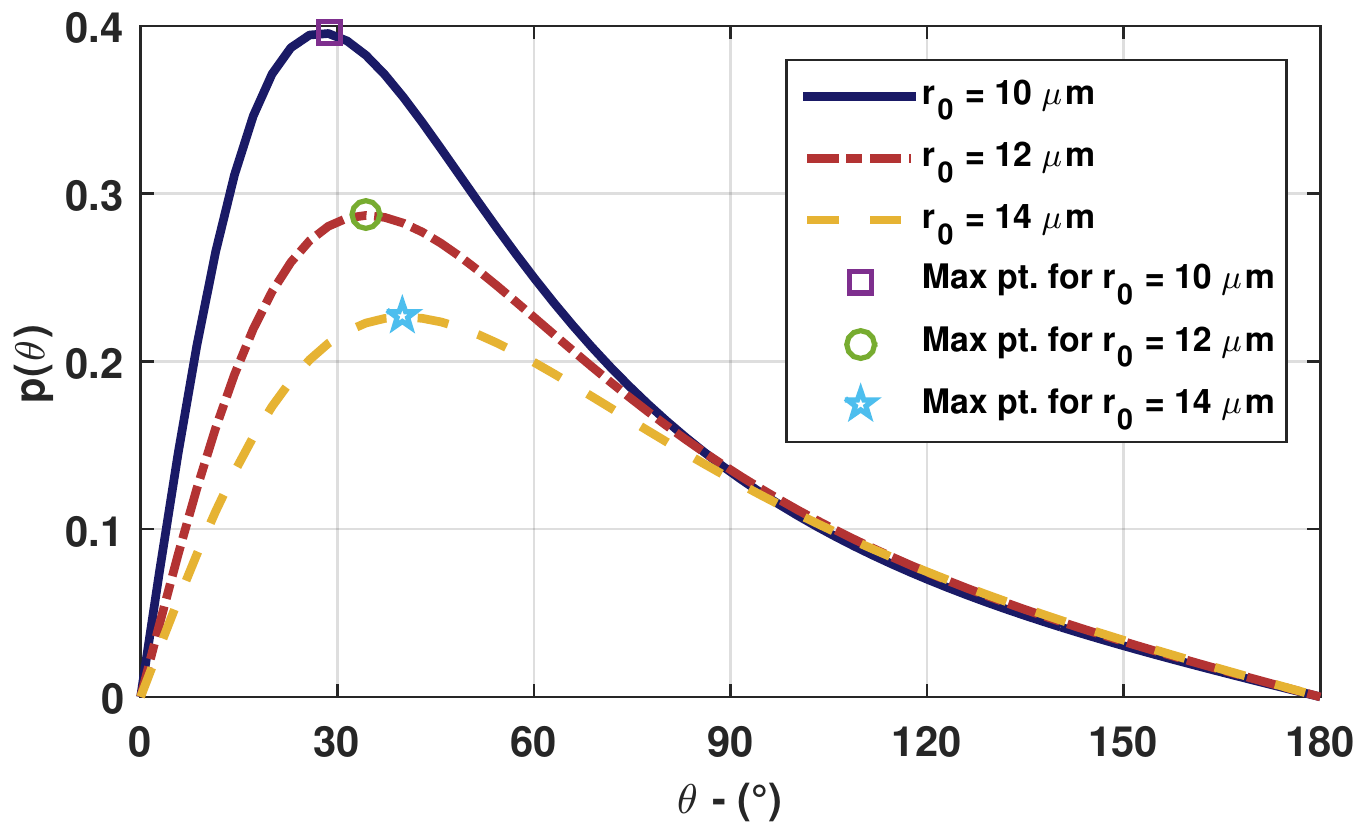}
		\caption{Theta versus $\ptheta$ curves for different $r_0$ values (${r_r = \SI{5}{\micro\meter}}$). Maximum values are attained at $\SI{28.6}{\degree}$, $\SI{34.3}{\degree}$, and $\SI{40.1}{\degree}$.}
		\label{fig_ptheta}
	\end{center}
\end{figure}

In particular, $\ptheta$ in  \eqref{eq:teta} gives the distribution of the molecules absorbed by the sphere with respect to angle $\theta$ which is defined in Fig. \ref{fig_fig1}. This function is plotted for different parameters in Fig. \ref{fig_ptheta}. As can be seen from this figure, the probability of absorption has a peak between $\theta$=\SI{0}{\degree} and $\theta$=\SI{90}{\degree}. Furthermore, it is zero for $\theta$ = \SI{0}{\degree} and $\theta$ =\SI{180}{\degree}. These are not surprising since  $\theta$=\SI{0}{\degree} and  $\theta$=\SI{180}{\degree} represent only a point on the surface. Therefore, the probability of absorption in these regions is zero although  $\theta$=\SI{0}{\degree} is the closest point to the transmitter. If we increase $\theta$, we expect to have more received molecules since the circular region gets bigger. However, after some point, the rate of increase is not enough compared to the decrease in the hitting rate that can also be observed in Fig.~\ref{fig_ptheta}. 


\tbayram{Since the communication process occurs in a limited time, we need to obtain the joint distribution of absorbed molecules with respect to time and angle to apply partially counting receiver system in MC.} To the best of our knowledge, the joint distribution with respect to time and angle has not been derived yet. By utilizing \eqref{fhit} and \eqref{eq:teta}, we find an approximate analytical closed-form expression for the joint cumulative distribution with respect to time and angle. The main concept of our approach is to cover the desired region on the surface of the spherical receiver with infinitesimally small spheres and to evaluate the absorption probability of these spheres. We also add compensation functions in multiplication form and solve them by utilizing \eqref{fhit} and \eqref{eq:teta}, i.e., the known marginal cases in the literature. 

When we consider an infinitesimally small sphere with radius $dr$ placed at the surface making an angle of $\theta\,$ with the center of the sphere as shown in  Fig.\ref{fig_fig1}, the distance of this arbitrarily placed sphere to the point transmitter can be calculated using Cosine rule as

\begin{equation}
{ r }_{ 0 }^{ * }=\sqrt { { r }_{ 0 }^{ 2 }+{ r }_{ r }^{ 2 }-2{ r }_{ 0 }{ r }_{ r }\cos { \theta  }  }. 
\label{eq:r0}
\end{equation}

Considering Fig. \ref{fig_fig1}, if only the small sphere at point $\pointA$ is available in the environment, the probability of absorption of molecules until time $t$ could be obtained using \eqref{fhit} as
\begin{equation}
\fhitA{\pointA} = \fhit{r_0^*}{dr} = \cfrac { dr }{ { r }_{ 0 }^{ * } }\, \merfc {\cfrac { { r }_{ 0 }^{ * }-dr }{ \sqrt { 4Dt }  }  }. 
\label{fhit2}
\end{equation}
Since ${ r }_{ 0 }^{ * } \gg  dr$, we have ${ { r }_{ 0 }^{ * }-dr } \approx  { r }_{ 0 }^{ * }$ and, therefore, \eqref{fhit2} can be rewritten as 

\begin{equation}
\fhitA{\pointA} \approx \cfrac { dr }{ { r }_{ 0 }^{ * } }\, \merfc {\cfrac { { r }_{ 0 }^{ * } }{ \sqrt { 4Dt }  }  }.
\label{fhit22}
\end{equation}

\tbayram{When the big sphere also acts as another absorber, some part of the small sphere lies behind the surface of the big sphere; hence this part does not act as a receiver. Let $\pointApr$ be the region of the active receptors of the small sphere placed at point $\pointA$ of the big sphere whose receptors are also active. When the big sphere acts as a receiver, the small sphere at point $\pointA$ can only absorb a molecule unless it has not been absorbed by the big sphere earlier. Therefore,  we add $(1-{ F }_\text{ hit }(t)+{ F }_\text{ hit }^A (t))$ as a factor to $\fhitA{\pointA}$ in order to model this event. Furthermore, an angle factor $\afactor$ and a time factor $\tfactor$ are also added to $\fhitA{\pointA}$ as the adjusting factors, which will be derived using marginal cases. Since the orientation of the active regions of the small spheres are different for different $\theta$, these adjusting factors are necessary. Combining all of these arguments, the probability of absorption of a molecule until time $t$ with the active regions of the small sphere at point $\pointA$, $\fhitA{\pointApr}$, can be approximated as}
\begin{align}
\begin{split}
\fhitA{\pointApr} &= \![1 \!-\! \fhit{r_0}{r_r}+\fhitA{\pointA}] \, \fhitA{\pointA} \, \afactor \tfactor \\
          &\approx\! [1 \!- \! \fhit{r_0}{r_r}] \frac { dr }{ { r }_{ 0 }^{ * } }\, \merfc {\frac { { r }_{ 0 }^{ * } }{ \sqrt { 4Dt }  }  } \, \afactor \tfactor,
\end{split}
\label{fhit4}
\end{align}
since $\fhitA{\pointA} \ll \fhit{r_0}{r_r}$.

\begin{figure}[!t]
	\begin{center}
\includegraphics[width=0.62\columnwidth,keepaspectratio]%
		{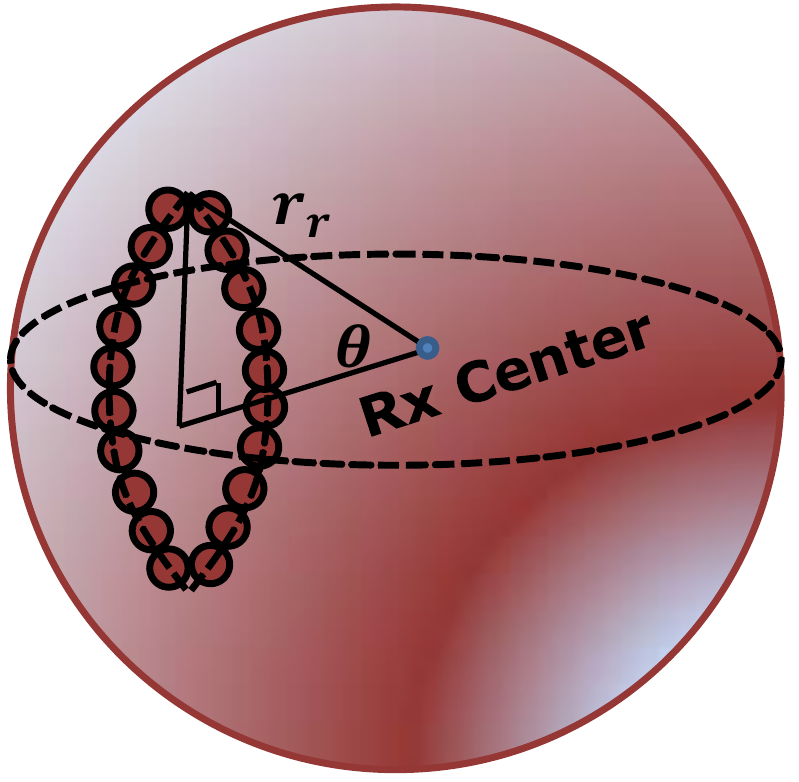}
		\caption{Small spheres that make $\theta$ angle with the center of the big sphere.}
		\label{fig_fig2}
	\end{center}
\end{figure}

After approximating $\fhitA{\pointApr}$, for a small sphere placed at point $\pointA$ that makes $\theta$ angle as shown in Fig.\ref{fig_fig1}, the next step is to find the total number of small spheres that have similar $\theta$ angle. These small spheres are lined up on a circle with radius $r_r \sin(\theta)$, as shown in Fig. \ref{fig_fig2}. Since the radius of these spheres are infinitesimal, the total number of spheres on this circle can be calculated by dividing the circumference of the circle to the diameter of the small sphere as $\ntheta=\cfrac { 2\pi { r }_{ r }\sin(\theta)  }{ 2dr }$. Using this $\ntheta$, the probability of absorption of a molecule until time $t$ by any small sphere that makes same $\theta$ angle as the sphere at point A can be obtained by 

\begin{align}
\begin{split}
&\pthetaT{t} = \ntheta \fhitA{\pointApr} \\
&= \pi r_r \sin(\theta) [1 \!- \! \fhit{r_0}{r_r}] \frac { \merfc {\frac { { r }_{ 0 }^{ * } }{ \sqrt { 4Dt }  }  } }{ { r }_{ 0 }^{ * } }\, \afactor \tfactor .
\label{fhit5}
\end{split}
\end{align}
When $t$ goes to infinity, \eqref{fhit5} can be equalized to \eqref{eq:teta}, where  $\lim\limits_{t\rightarrow \infty}p(\theta,t)$ can be obtained as
\begin{equation}
\lim_{t\rightarrow \infty}\pthetaT{t} =  \pi { r }_{ r }\sin { \theta  } (1-\frac {r_r}{r_0}) \frac { 1 }{ r_0^* }  \afactor \lim_{t\rightarrow \infty}\tfactort{t}.
\label{fhit6}
\end{equation}
Hence, equalizing \eqref{eq:teta} and \eqref{fhit6} gives us $\afactor$ as
\begin{equation}
\afactor = \frac { 2 r_r r_0^* \,\, \epstheta }{ \left( 1-\frac { r_r }{ r_0 }  \right) \lim\limits_{t\rightarrow \infty}\tfactort{t}  }. 
\label{Tfunc}
\end{equation}
Although \eqref{Tfunc} contains $\lim\limits_{t\rightarrow \infty}\tfactort{t}$, in the following steps this term is canceled out and $\pthetaT{t}$ does not involve any limit term.

The next step is deriving the other compensation function, $\tfactor$. Note that $\pthetaT{t}$ gives the distribution of molecules with respect to angle $\theta$ until time $t$. Therefore, taking the integral of $\pthetaT{t}$ with respect to $\theta$ from $\theta\!=\!0$ to an arbitrary angle $\alpha$, gives the cumulative distribution of molecules at the receiver with respect to time and angle as 
\begin{equation}
F\left(\alpha, t\right) =\int _{ 0 }^{\alpha} {\pthetaT{t} \, d\theta}. 
\label{cum}
\end{equation}
In \eqref{cum}, one can easily observe that, when $\alpha=\pi$, all of the surface of the receiver is absorbing. Therefore, $F\left( \pi ,t \right)$ is equal to the marginal cumulative function given in \eqref{fhit} as $F\left( \pi ,t \right)= \fhit{r_0}{r_r}$. By using this equality, we can obtain $\tfactor$ as 
\begin{align}
\begin{split}
\tfactor = \frac{\fhit{r_0}{r_r}}{ \pi r_r [1 \!- \! \fhit{r_0}{r_r}] S_{\pi} },
\end{split}
\label{Afunc}
\end{align}
where
\begin{align}
S_{\pi} &= \int_0^{\pi} \frac{\sin\theta}{r_0^*} \merfc{ \frac{r_0^*}{\sqrt{4Dt}} } \afactor \, d\theta .
\end{align}
Note that the denominator of $\tfactor$ contains $\afactor$. Since $\lim\limits_{t\rightarrow \infty}\tfactort{t}$ term in this integral can be taken outside of the integral, we conclude that $\tfactor$ involves $\lim\limits_{t\rightarrow \infty}\tfactort{t}$ in the numerator while $\afactor$ involves this term in the denominator. Therefore, multiplying these two compensation functions together cancels  $\lim\limits_{t\rightarrow \infty}\tfactort{t}$ terms in $\pthetaT{t}$.

After finding $\tfactor$, we can write $\pthetaT{t}$ as
\begin{align}
\pthetaT{t} = \frac{ \sin\theta \,\merfc{\frac{r_0^*}{\sqrt{4Dt}}}  \fhit{r_0}{r_r} }{(1\!\!-\!\!\frac{2r_r}{r_0}\cos\theta \!\!+\! \frac{r_r^2}{r_0^2})^{\frac{3}{2}} \bigintss_0^{\pi} \!\frac{\sin\theta' \,\merfc{\frac{r_0^*}{\sqrt{4Dt}}}}{(1\!-\frac{2r_r}{r_0}\cos\theta' \!+ \frac{r_r^2}{r_0^2})^{\frac{3}{2}}}  \,d\theta'}.
\label{eq:floatingeq}
\end{align}

\begin{figure}[t] 
\includegraphics[width=0.9\columnwidth,keepaspectratio]{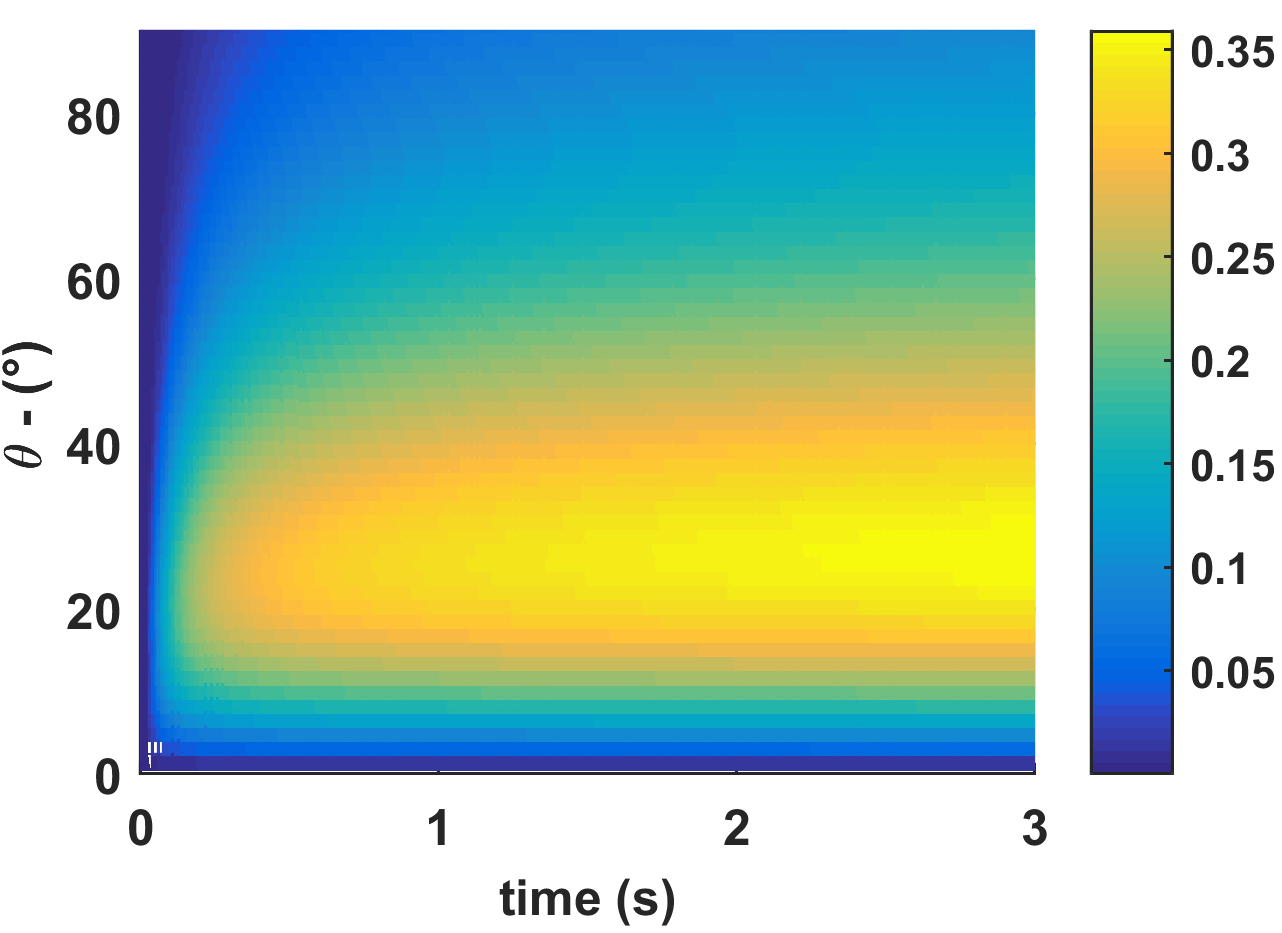}
\caption{$\pthetaT{t}$ heat map  for $r_r$=5 $\mu m$, $r_0$=10 $\mu m$, and $D=80 { \mu { m }^{ 2 } }/{ s }$. }
 \label{fig:pteta}
\end{figure}

{Fig. \ref{fig:pteta} is the heat map of $\pthetaT{t}$ that gives the angular distribution of a molecule until time $t$. Considering this figure, some interesting inferences can be obtained. Firstly, the molecules will accumulate less at the higher angles compared to the lower angles. This is expected since as the angle increases, the distance also increases, which leads to the diminishing of the probability of absorption. Secondly and more interestingly, at very small angles around zero that can also be considered as the line of sight angles, the probability of absorption is even lower compared to other angles. This is a consequence of the fact that the number of small spheres is very limited for these angles (when $\theta$ = \SI{0}{\degree}, there is only one small sphere); hence, the probability of absorption at these angles is quite low.}


Note that $\pthetaT{t}$ is used for calculating $F\left( \alpha ,t \right)$ as in  \eqref{eq:floatingeq2} where $Ei(.)$ is an exponential integral function.

%
\addtocounter{equation}{1}%
\setcounter{storeeqcounter}%
{\value{equation}}%

\begin{figure*}[!t]
\normalsize
\setcounter{tempeqcounter}{\value{equation}} 
\begin{IEEEeqnarray}{rCl}
\setcounter{equation}{\value{storeeqcounter}} 
\begin{aligned}
& F\left( \alpha ,t \right)=\frac { { \erfc }\left( \frac { (r_{ 0 }-{ r }_{ r }) }{ \sqrt { 4Dt }  }  \right) \left( Dt\quad { \erfc }\left( \frac { \sqrt { r_{ 0 }^{ 2 }-2r_{ 0 }{ r }_{ r }+{ r }_{ r }^{ 2 } }  }{ \sqrt { 4Dt }  }  \right) +\frac{1}{\sqrt { 2\pi  } } \sqrt { Dt } \sqrt { r_{ 0 }^{ 2 }-2r_{ 0 }{ r }_{ r }+{ r }_{ r }^{ 2 } } { Ei }\left( -\frac { \left( r_{ 0 }^{ 2 }-2r_{ 0 }{ r }_{ r }+{ r }_{ r }^{ 2 } \right)  }{ 4Dt }  \right)  \right)  }{ U(t)\quad Dt\sqrt { \frac { r_{ 0 }^{ 2 }-2r_{ 0 }{ r }_{ r }+{ r }_{ r }^{ 2 } }{ r_{ 0 }^{ 2 } }  }  } \\
&-\frac { { \erfc }\left( \frac { (r_{ 0 }-{ r }_{ r }) }{ \sqrt { 4Dt }  }  \right) \left( Dt { \erfc }\left( \frac { \sqrt { r_{ 0 }^{ 2 }-2r_{ 0 }{ r }_{ r }\cos  (\alpha )+{ r }_{ r }^{ 2 } }  }{ \sqrt { 4Dt }  }  \right) +\frac{1}{\sqrt { 2\pi  } } \sqrt { Dt } \sqrt { r_{ 0 }^{ 2 }-2r_{ 0 }{ r }_{ r }\cos  (\alpha )+{ r }_{ r }^{ 2 } } { Ei }\left( -\frac { \left( r_{ 0 }^{ 2 }-2r_{ 0 }{ r }_{ r }\cos  (\alpha )+{ r }_{ r }^{ 2 } \right)  }{ 4Dt }  \right)  \right)  }{ U(t)\quad Dt\sqrt { \frac { r_{ 0 }^{ 2 }-2r_{ 0 }{ r }_{ r }\cos  (\alpha )+{ r }_{ r }^{ 2 } }{ r_{ 0 }^{ 2 } }  }  } 
\end{aligned}
\label{eq:floatingeq2}
\end{IEEEeqnarray}
\setcounter{equation}{\value{tempeqcounter}} 
\hrulefill
\vspace*{4pt}
\end{figure*}

%
\addtocounter{equation}{1}%
\setcounter{storeeqcounter}%
{\value{equation}}%

\begin{figure*}[!t]
\normalsize
\setcounter{tempeqcounter}{\value{equation}} 
\begin{IEEEeqnarray}{rCl}
\setcounter{equation}{\value{storeeqcounter}} 
\begin{aligned}
& U(t)=  { \displaystyle{ \int _{ 0 }^{ \pi  }}{ \erfc({ \cfrac { \sqrt { { r }_{ 0 }^{ 2 }+{ r }_{ r }^{ 2 }-2{ r }_{ 0 }{ r }_{ r }\cos { \theta  }  }  }{ \sqrt { 4Dt }  } ) }\quad \cfrac { \sin { \theta  }  }{ { \left( 1-\cfrac { 2{ r }_{ r } }{ { r }_{ 0 } } \cos { \theta  } +\cfrac { { r }_{ r }^{ 2 } }{ { r }_{ 0 }^{ 2 } }  \right)  }^{ { 3 }/{ 2 } } } \quad d\theta }} = \\
& \frac { { { r }_{ 0 } }^{ 2 }\left( Dt\quad { \erfc }\left( \frac { { r }_{ 0 }-{ r }_{ r } }{ \sqrt { 4Dt }  }  \right) +\sqrt { Dt } \left( { r }_{ 0 }-{ r }_{ r } \right) { Ei }\left( -\frac { { \left( { r }_{ 0 }-{ r }_{ r } \right)  }^{ 2 } }{ 4Dt }  \right)  \right)  }{ D\quad { r }_{ r }t\quad { (r }_{ 0 }-{ r }_{ r }) } -\frac { { { r }_{ 0 } }^{ 2 }\left( Dt\quad { \erfc }\left( \frac { { r }_{ 0 }{ +r }_{ r } }{ \sqrt { 4Dt }  }  \right) +\sqrt { Dt } \left( { r }_{ 0 }{ +r }_{ r } \right) { Ei }\left( -\frac { { \left( { r }_{ 0 }{ +r }_{ r } \right)  }^{ 2 } }{ 4Dt }  \right)  \right)  }{ D{ r }_{ r }t\quad { (r }_{ 0 }{ +r }_{ r }) } 
\end{aligned}
\label{eq:floatinge22}
\end{IEEEeqnarray}
\setcounter{equation}{\value{tempeqcounter}} 
\hrulefill
\vspace*{4pt}
\end{figure*}

Once $F\left( \alpha ,t \right)$ is obtained, the channel tap for the $\xth{n}$ symbol duration, $p_n$, can be obtained (for a given counting region that is defined by $\alpha$) as
\begin{equation}
p_n(\alpha)=F\left( \alpha ,n t_{s} \right)-F\left( \alpha ,(n\!-\!1)t_{s} \right).
\label{cum3_maybe_multiple_defn}
\end{equation}

\section{Channel Model Validation and Molecular Signal Properties\label{sec_signal_properties}}

\begin{figure*}[th]
	\subfloat[$r_r=5 \mu m$, $r_0$=10  $\mu m$ and $D=80 { \mu { m }^{ 2 } }/{ s }$]{%
\includegraphics[width=0.48\textwidth]{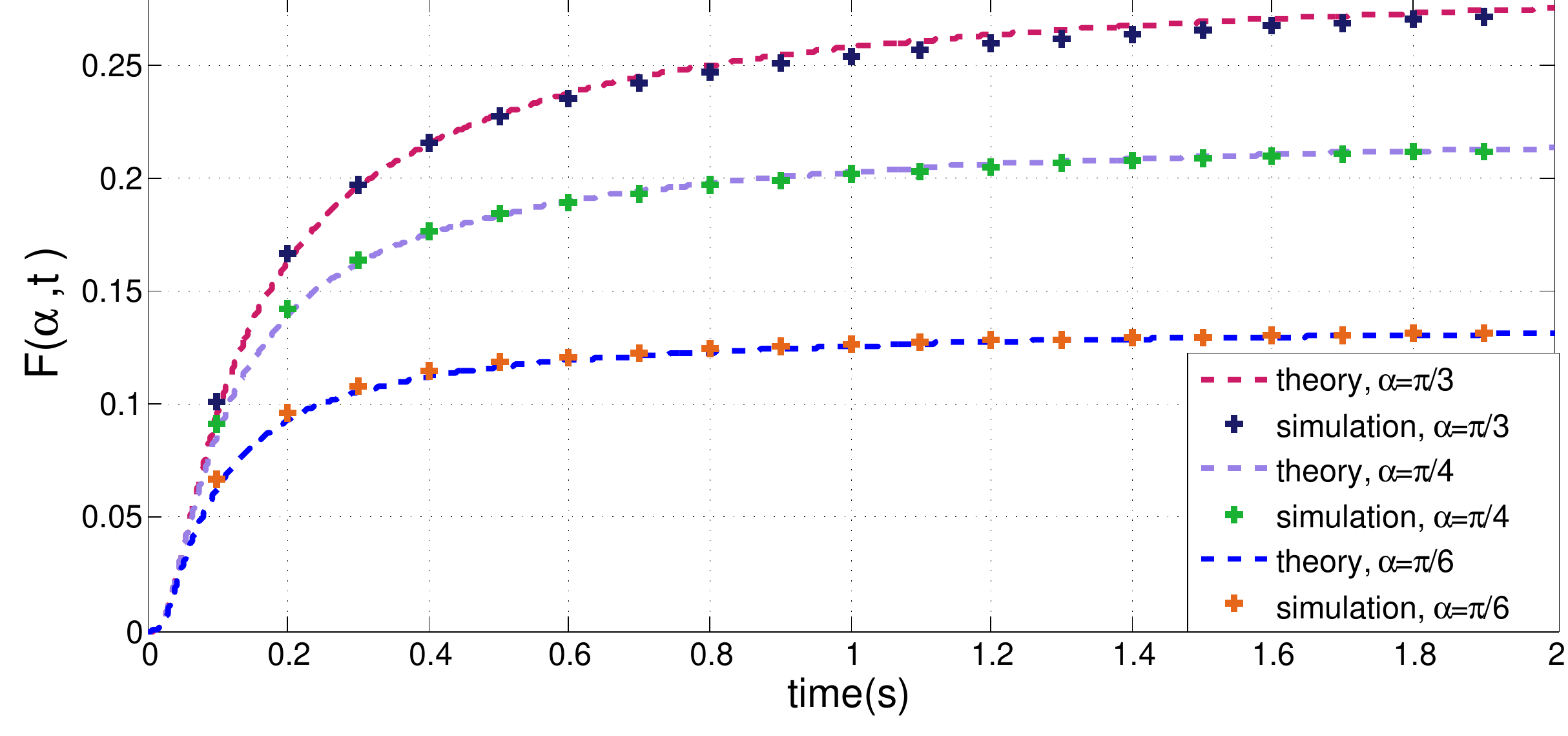}}
    \label{fig:ap}\hfill
    \subfloat[$r_r=5 \mu m$, $r_0=9 \mu m$ and $D=80 { \mu { m }^{ 2 } }/{ s }$]{%
    \includegraphics[width=0.48\textwidth]{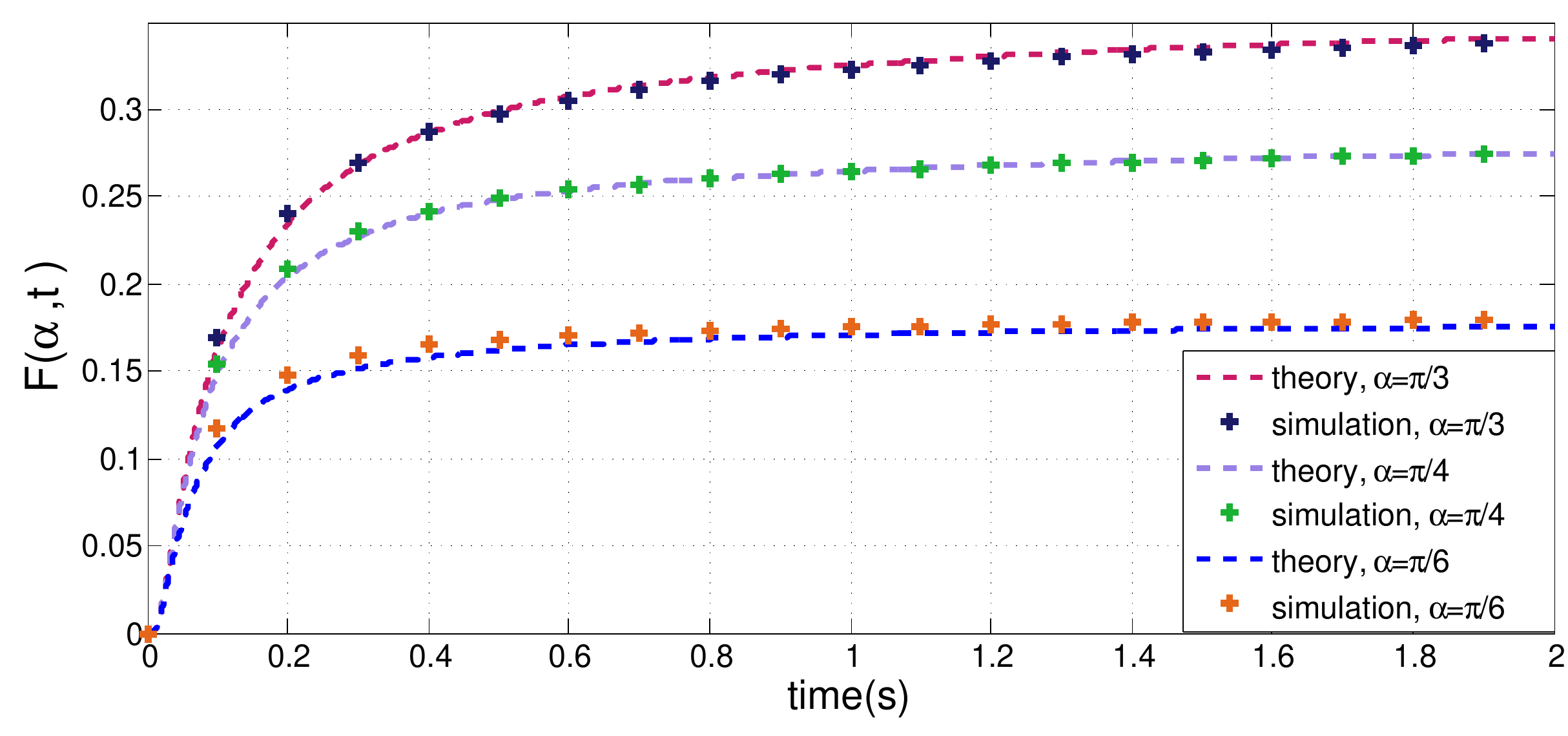}}
    \label{fig:e}\\
    \subfloat[$r_r$=5 $\mu m$, $r_0=10 \mu m$ and $D=160{ \mu { m }^{ 2 } }/{ s }$]{%
    \includegraphics[width=0.48\textwidth]{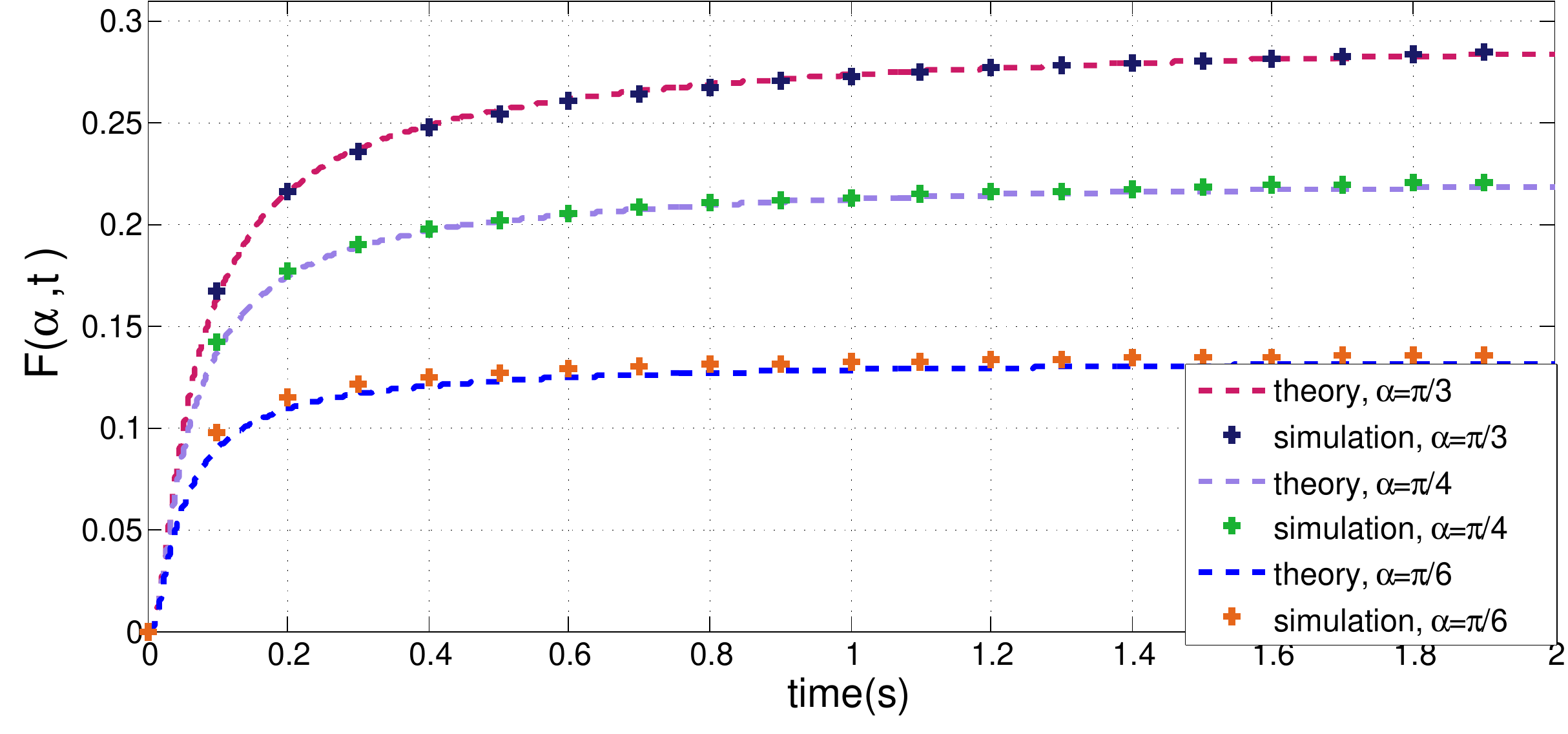}}
    \label{fig:c}\hfill
    \subfloat[$r_r$=10 $\mu m$, $r_0$=20  $\mu m$ and $D=80{ \mu { m }^{ 2 } }/{ s }$]{%
    \includegraphics[width=0.48\textwidth]{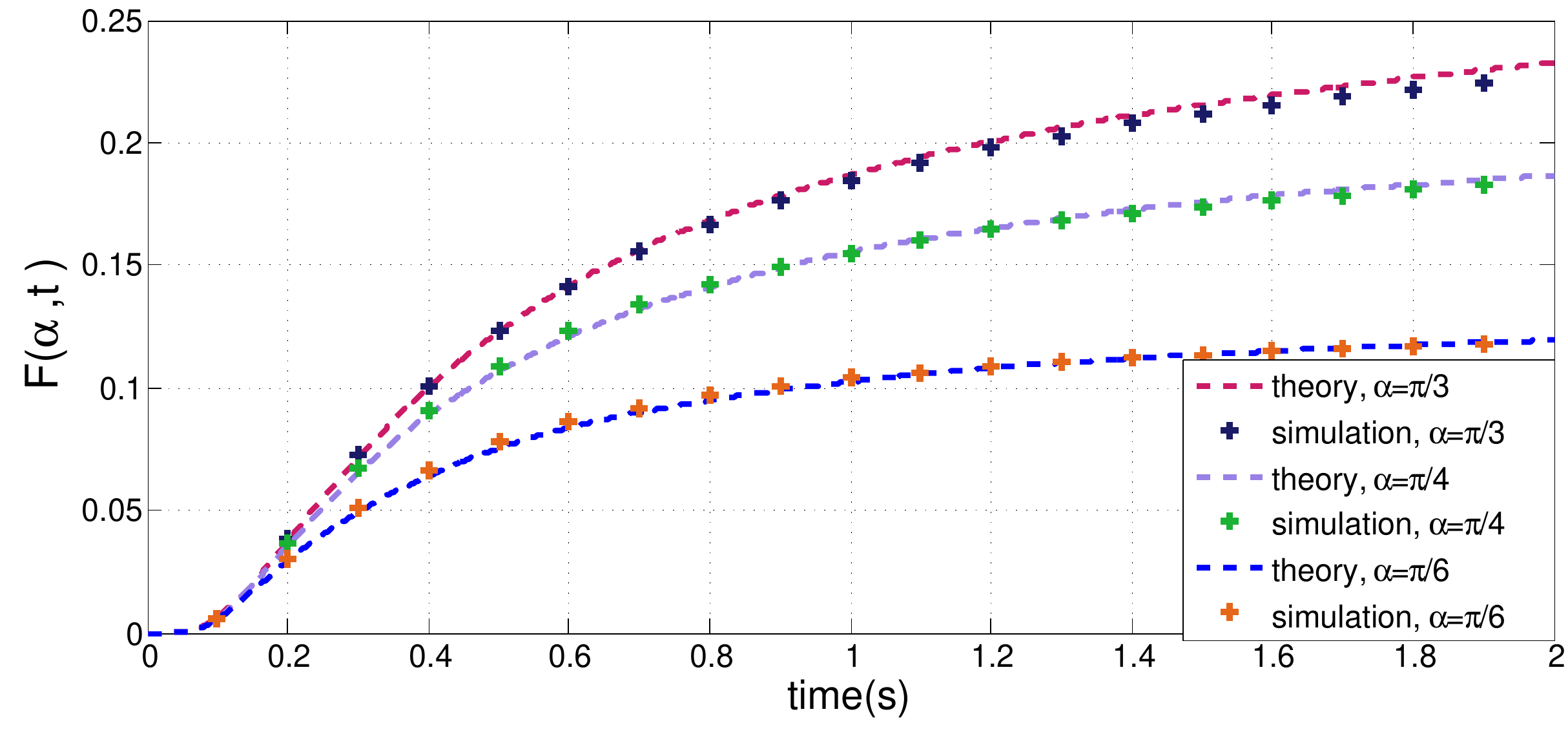}}
    \label{fig:d}
\caption{Comparison of the analytical cumulative function obtained in Eq. (\ref{eq:floatingeq2}) with simulation results for $\alpha$= $\pi/3$, $\pi/4$  and  $\pi/6$ from top to bottom } \label{fig:validation}
\end{figure*}

\subsection{Received Signal Validation}
Once the analytical distribution of the received molecules for partially counting system are obtained, the next step is \tbirkan{to compare it} with the simulation results obtained by using \eqref{brown}. As can be seen in Fig. \ref{fig:validation}, validation is done for various parameters with different $\alpha$ values, and simulation and theoretical results are coherent.

\subsection{Peak Time}
\begin{figure}[ht!] 
\includegraphics[width=0.99\columnwidth,keepaspectratio]{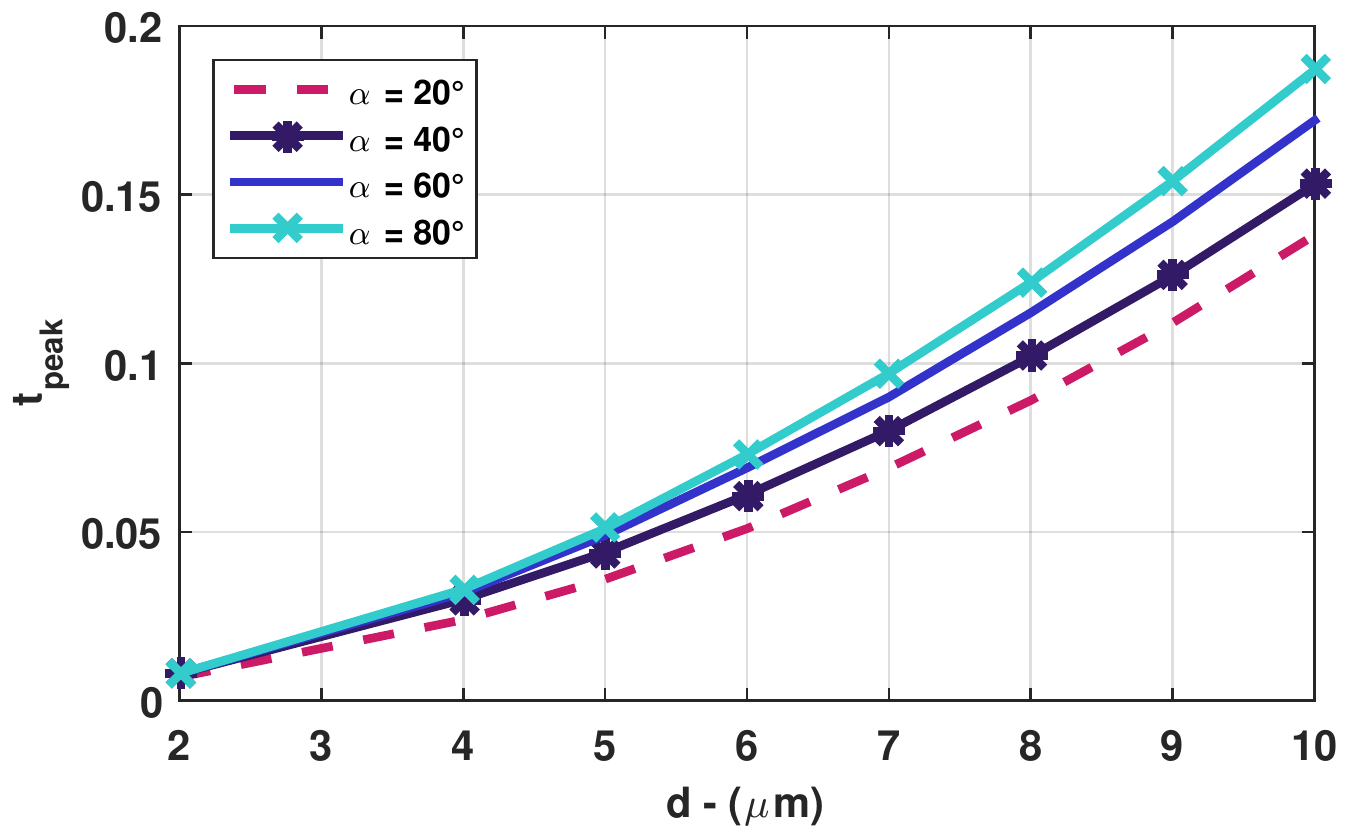}
\caption{$t_{peak} vs d$ curves  for $r_r=5 \mu m$, $r_0=10 \mu m$,  $D=80 { \mu { m }^{ 2 } }/{ s }$. }
 \label{fig:peak2}
\end{figure}

{The communication literature considers the peak time, $t_{peak}$, to be a crucial property for characterizing the channel, and defines it as the time that the received signal makes a peak at the receiver. In \cite{yilmaz2014threeDC}, it is concluded that, for the fully absorbing receiver, $t_{peak}$ is proportional with  the square of $d=r_0-r_r$, which is the shortest distance from the transmitter to the receiver's surface. This is a major drawback in molecular communication via diffusion (MCvD) channels since as $d$ increases, the data rate exponentially decreases to capture the signal until its peak, while in electro-magnetic communication this decrement is linear. We evaluate $t_{peak}$ by taking the derivative of $F\left( \alpha ,t \right)$ with respect to time, which is the hitting rate of the molecules for a given $\alpha$, and examine its maximum value. As can be seen in Fig.~\ref{fig:peak2}, $t_{peak}$ is still directly proportional with $d^2$, which is the same with the fully absorbing receiver case.}

\subsection{ Optimum $\alpha$ for the Given Channel Parameters}
\begin{figure}[th] 
\includegraphics[width=0.95\columnwidth,keepaspectratio]{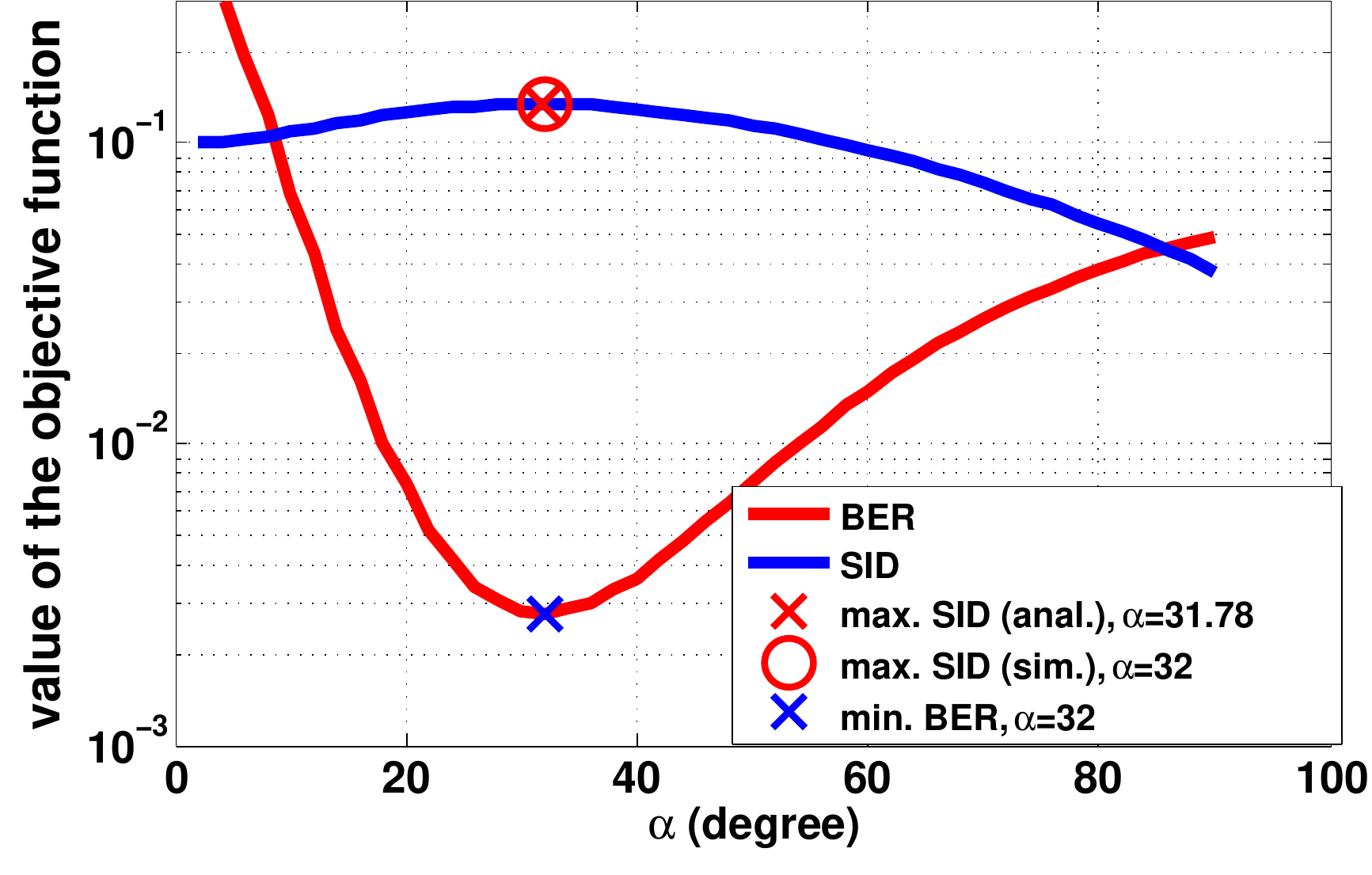}
\caption{BER vs $\alpha$ curves and corresponding SID curves for $r_r=5\mu m$, $r_0=10 \mu m$,  $D=80 { \mu { m }^{ 2 } }/{ s }$, $t_s=150ms$  with minimum point of BER function obtained via computer simulations and maximum of the SID function obtained with both  simulation and analytical solution of (\ref{eq:optAlfa}).}
 \label{fig:sid}
\end{figure}

The optimum reception angle, $\optangle$, of the receiver in terms of bit error rate (BER) is determined by finding the position of the global minimum of BER formula of the CSK modulation with respect to $\alpha$. On the other hand, closed-form of the BER formula in CSK is not a tractable function if  the number of the channel taps is high. Therefore, we  use an alternative objective function whose argument of the global maximum is almost the same as the argument of the global minimum of BER as proposed in \cite{pusane2017optimal}. This function is named as the signal to interference difference ($SID$), and gives the difference between the first tap and the sum of the other taps:
\begin{align}
\text{SID} = { {p_{1}(\alpha)}-{\sum_{n=2}^{\infty} p_{k}(\alpha)}} . 
\end{align}

As shown in Fig.~\ref{fig:sid}, the argument of the global maximum of this function is very close to the argument of the global minimum of BER. Using $SID$, the corresponding optimization problem is written as
\begin{align}
\optangle= \mathop{\mbox{arg max}}\limits_{ 0\le \alpha \le \pi } \left[{ {p_{1}(\alpha)}-{\sum_{k=2}^{\infty} p_{k}(\alpha)} }  \right].  
\label{eq:max11}
\end{align}
Since ${\sum_{k=2}^{\infty} p_{k}(\alpha)}= F\left( \alpha ,\infty \right)- p_{1}(\alpha)$, the optimization problem can be rewritten as
\begin{align}
\optangle= \mathop{\mbox{arg max}}\limits_{ 0\le \alpha \le \pi } \left[{ 2F\left( \alpha ,t_s \right) -F\left( \alpha ,\infty \right) }  \right].  
\label{eq:max111}
\end{align}

The solution of the optimization problem in \eqref{eq:max111} can be solved by taking the derivative of the objective function with respect to $\alpha$ and equating it to zero with reasonable simplifications. The corresponding solution can be obtained as 

\vspace{-5mm}

\begin{align}
\optangle=\cos ^{ -1 }{ \left( \frac { { r }_{ 0 }^{ 2 }+{ r }_{ r }^{ 2 }-{ \left( \cfrac { \sqrt { \pi a } (Y+M) }{ Y }  \right)  }^{ 2 }\quad  }{ 2{ r }_{ 0 }{ r }_{ r } }  \right)  }, 
\label{eq:optAlfa}
\end{align}
 where $a=D t_s$, $Y=\frac{-2 r_r \fhitx{r_0}{r_r}}{U(t_s)}$, and $M={ { r }_{ 0 }^{ 2 } }/{ 2 }$. The details of the derivation of $\optangle$ Ãis presented in Appendix.

\begin{figure*}[th] 
	\subfloat[$t_s$=100ms]{%
    \includegraphics[width=0.32\textwidth]{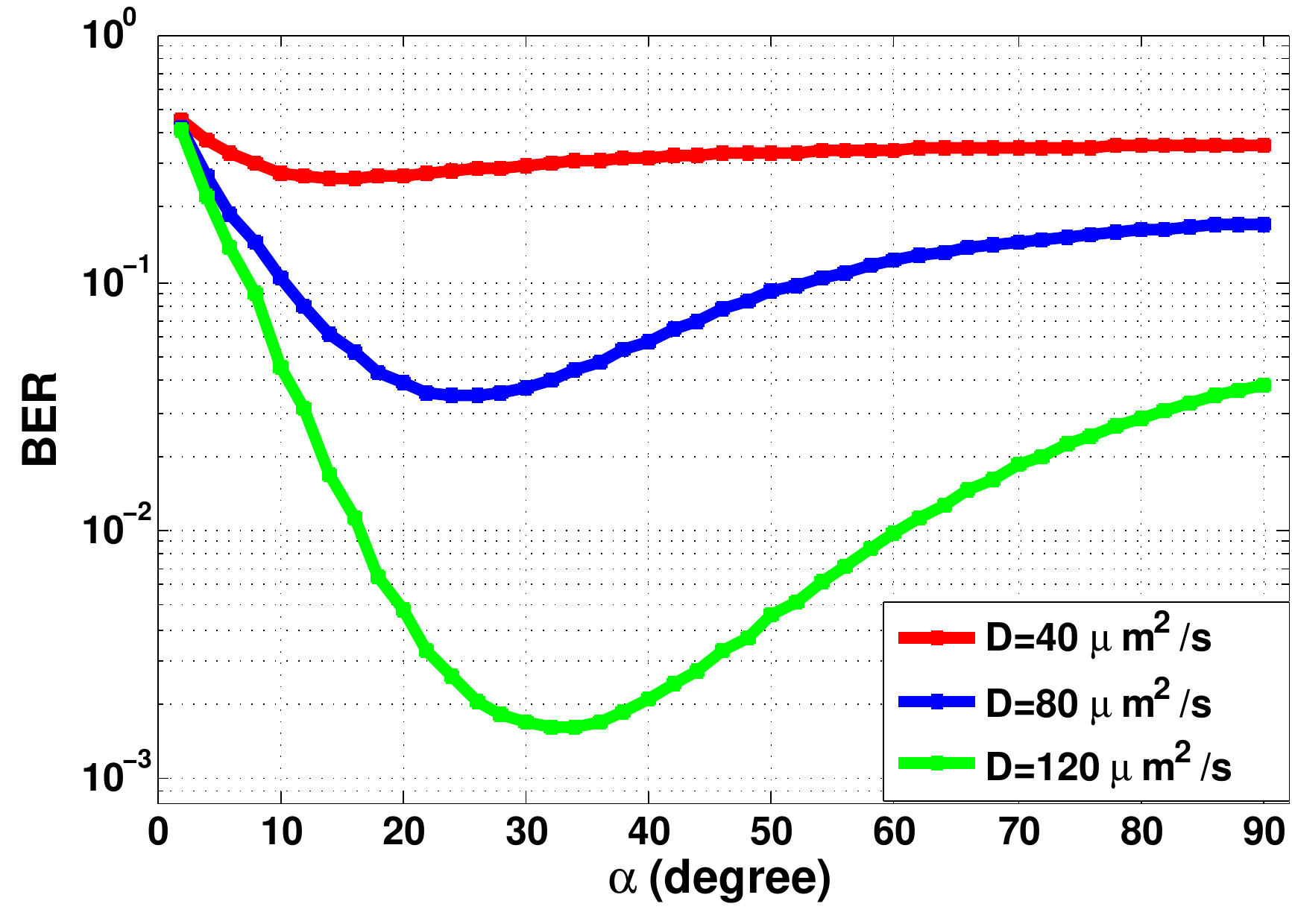}}
    \label{fig:ax2}\hfill
    \subfloat[$t_s$=150ms]{%
    \includegraphics[width=0.32\textwidth]{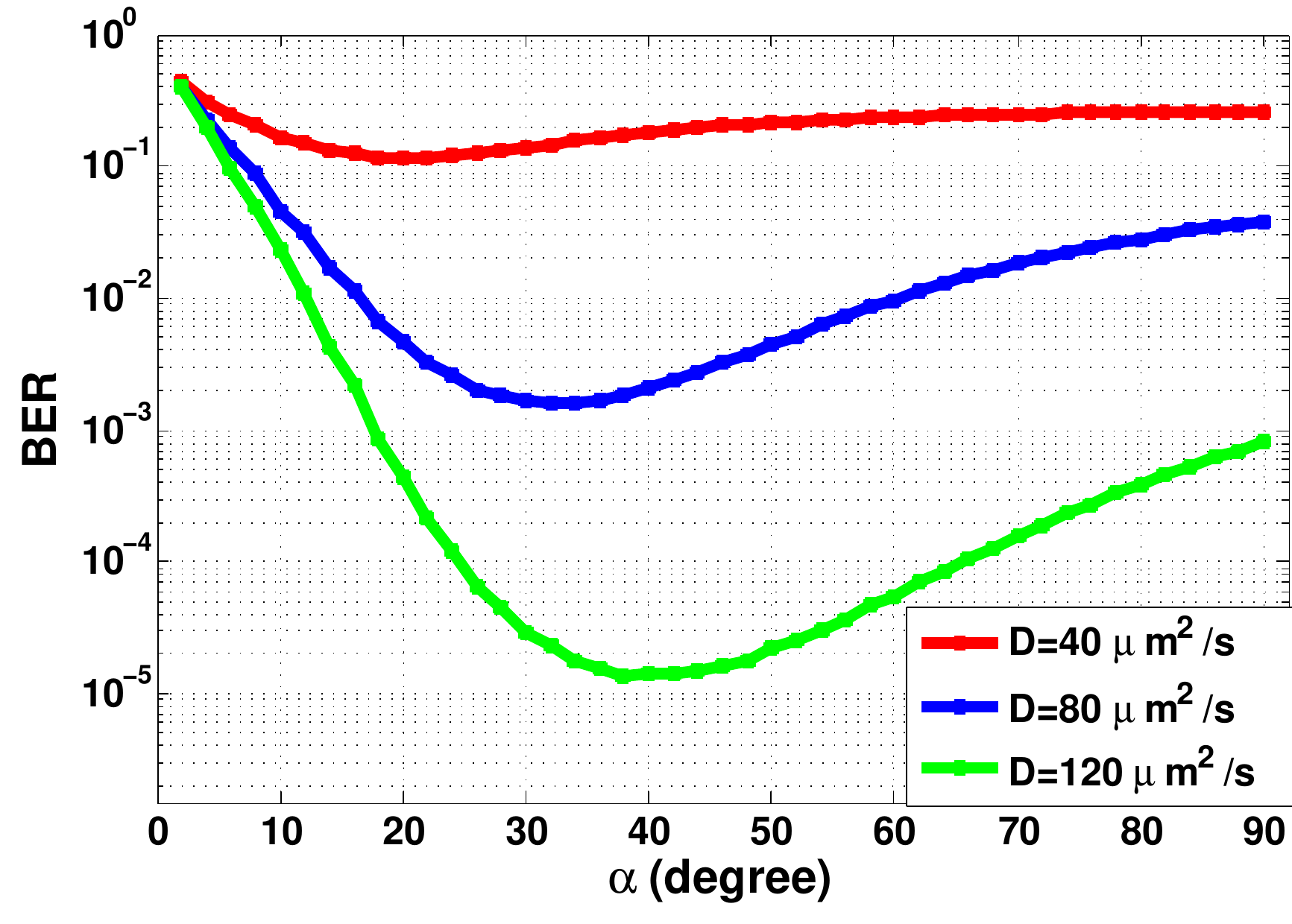}}
    \label{fig:bx2}\hfill 
    \subfloat[$t_s$=200ms]{%
    \includegraphics[width=0.32\textwidth]{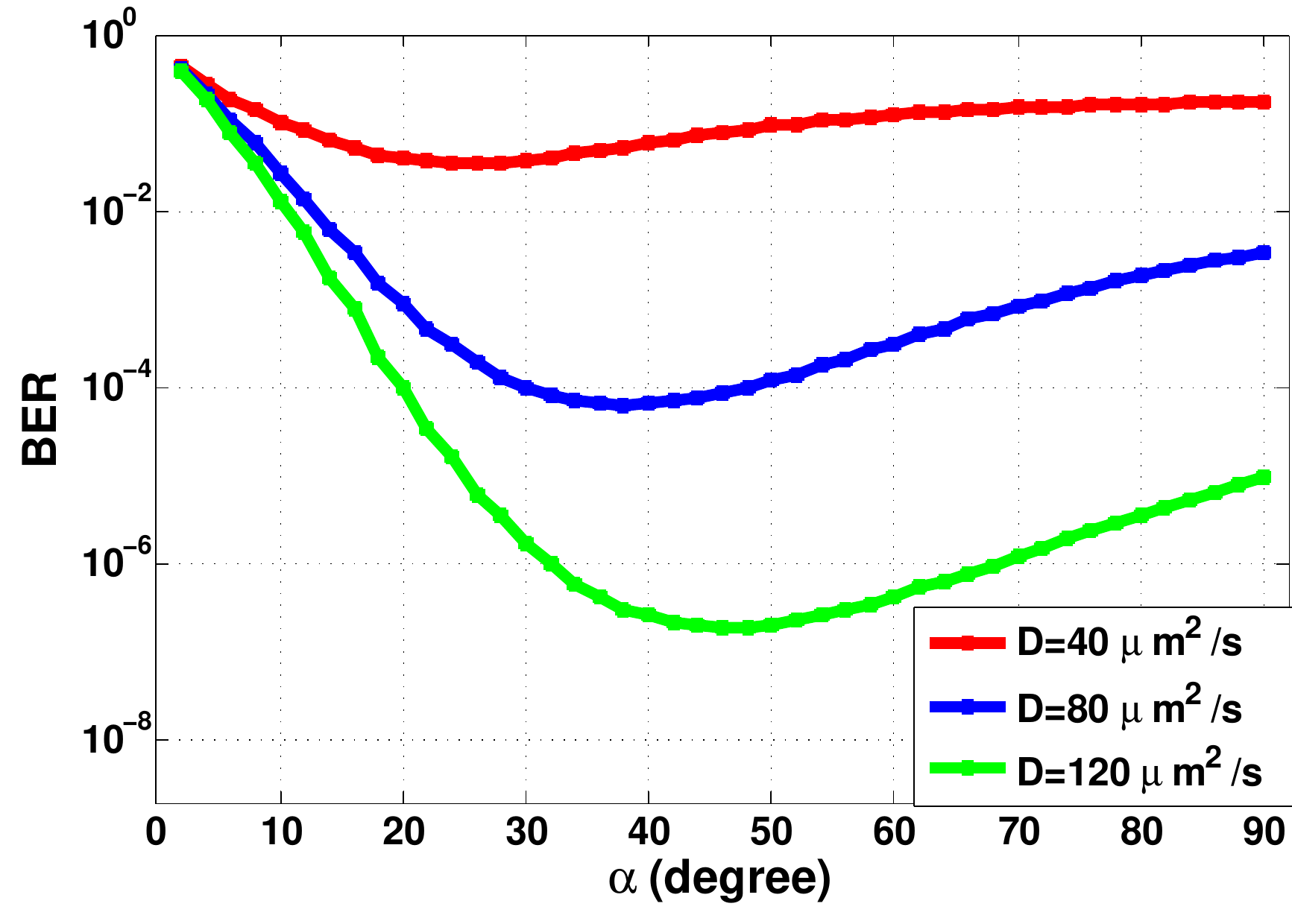}}
    \label{fig:bx22}
\caption{BER vs $\alpha$ curves for $r_r$=5 $\mu m$, for $r_0$=10  $\mu m$ with different diffusion coefficient ($D$) values}. 
\label{fig:differentD}
\end{figure*}

\section{Performance Analysis\label{sec_performance_analysis}}

In this section the performance analysis of the proposed system is examined for different parameters. We mainly evaluate the performance of the system in terms of BER. These evaluations are done using channel taps, both obtained  analytically by using \eqref{cum3_maybe_multiple_defn} and Monte Carlo simulations by releasing $10^5$ molecules from the transmitter and recording their arrival times and angles at the receiver. The channel taps are obtained using these records.

\begin{figure*}[th] 
	\subfloat[$t_s$=100ms]{%
    \includegraphics[width=0.32\textwidth]{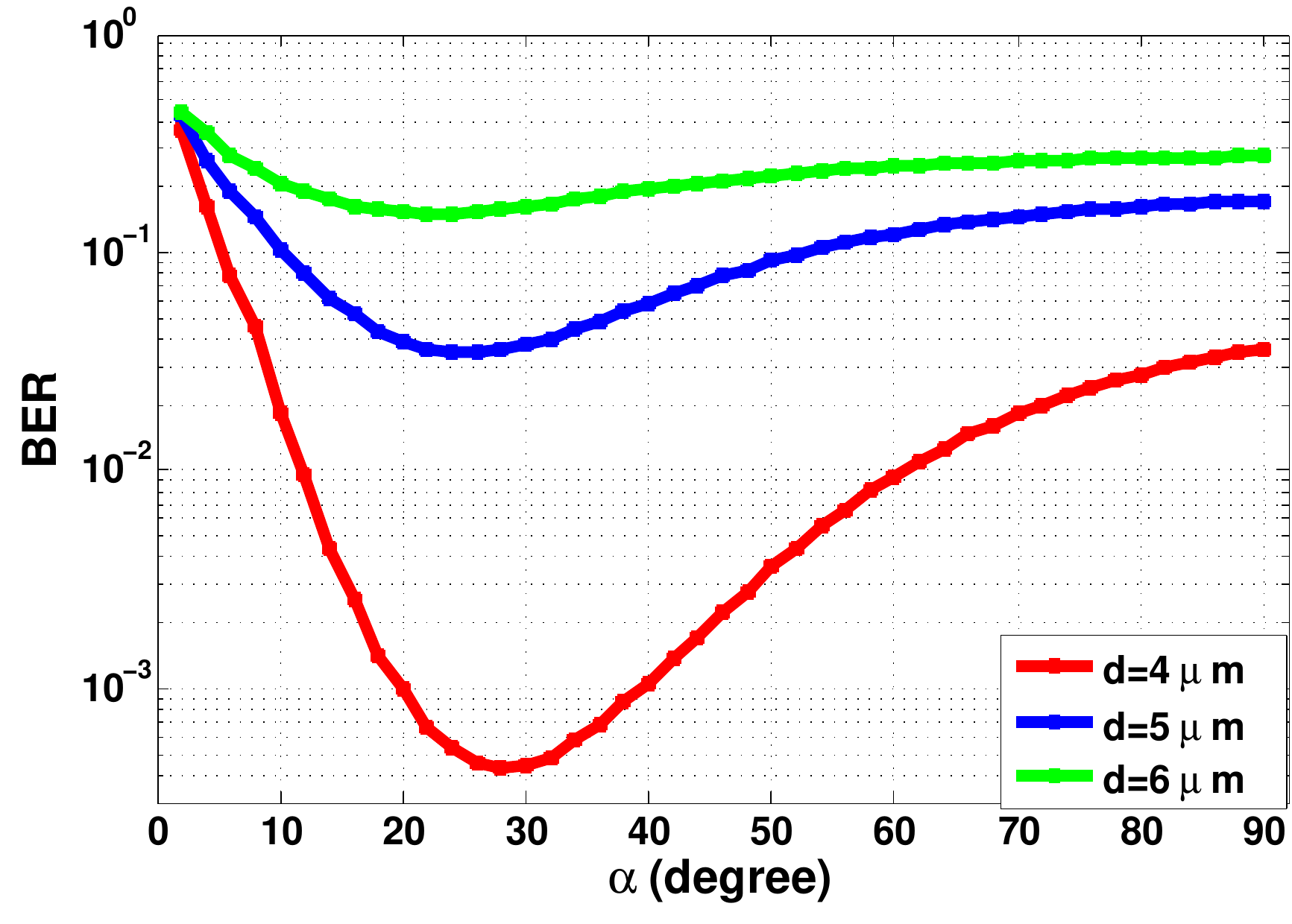}}
    \label{fig:ax2bb}\hfill
    \subfloat[$t_s$=150ms]{%
    \includegraphics[width=0.32\textwidth]{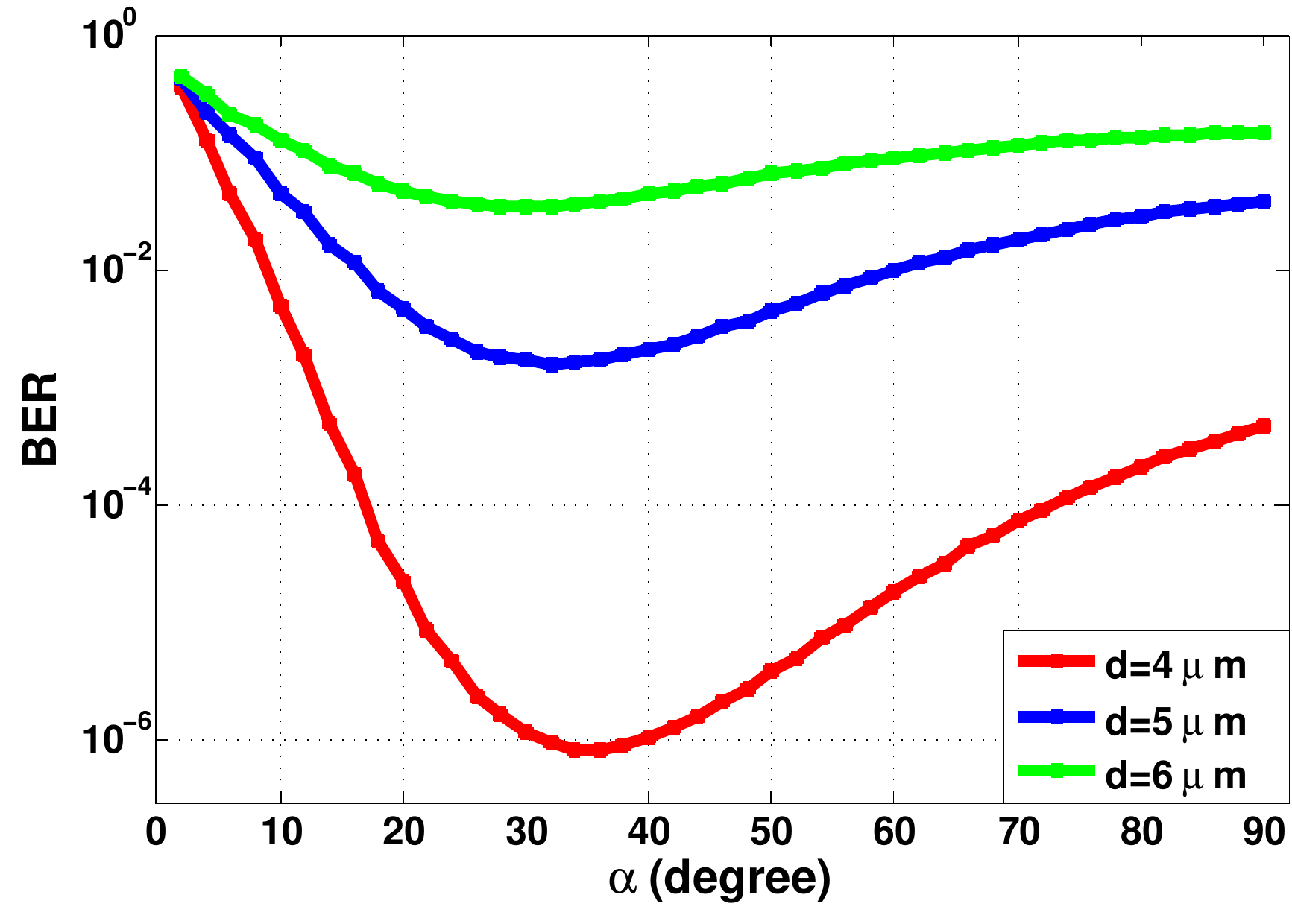}}
    \label{fig:bx2bb}\hfill 
    \subfloat[$t_s$=200ms]{%
    \includegraphics[width=0.32\textwidth]{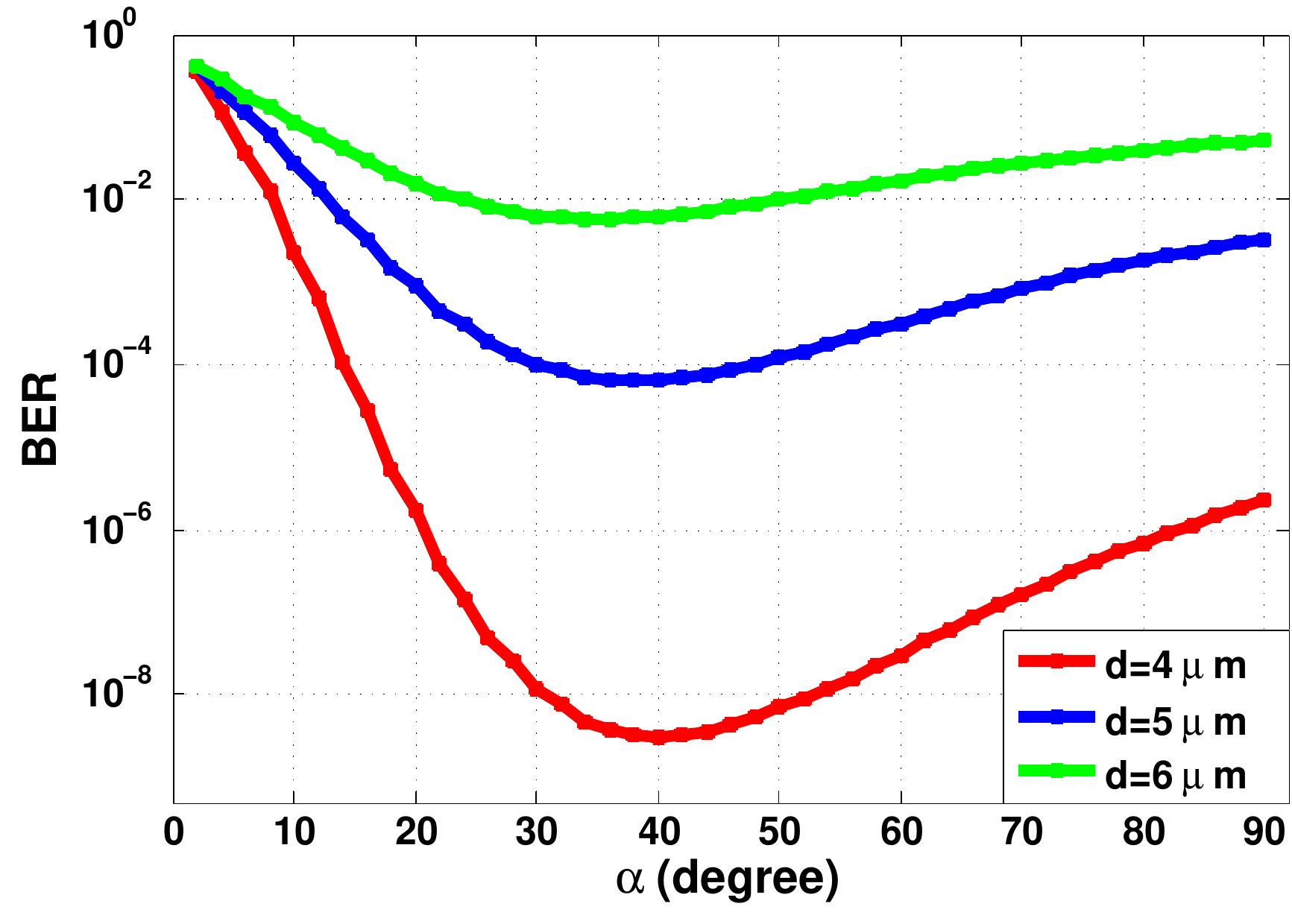}}
    \label{fig:bx22bb}
\caption{BER vs $\alpha$ curves for $r_r$=5 $\mu m$,  $D=80 { \mu { m }^{ 2 } }/{ s }$ with different $d$=$r_0$-$r_r$ values} 
\label{fig:differentR0}
\end{figure*}

We firstly evaluate the performance of the proposed system with respect to $d$ and the diffusion coefficient ($D$). As can be observed from Fig.~\ref{fig:differentD}, the optimum $\alpha$ in \tbirkan{terms of} BER increases as $D$ increases. This is expected since, as the molecules move faster, they can readily reach the further part of the receiver; hence, $\alpha$ should be increased in order not to miss the molecules coming \tbirkan{during the current} symbol \tbirkan{slot}. Similar results can be observed from  Fig.~\ref{fig:differentR0} where optimum $\alpha$ increases as the distance between the transmitter and the receiver decreases. Especially in the current time slot, the molecules can move towards the further parts of the spherical receiver as the distance decreases or $ D$ increases. Therefore, the relative gain of the first tap compared to other taps increases by increasing $\alpha$ when the distance is shorter or $D$ is higher. Furthermore, one can deduced from Fig. ~\ref{fig:differentD} and ~\ref{fig:differentR0}, as $t_s$ is increased the optimum $\alpha$ will also increases. This is also expected since optimum $\alpha$ will be  \SI{180}{\degree} when $t_s$ approaches to infinity.

\begin{figure*}[th] 
	\subfloat[BER curves for $t_s$=300ms and $d$=5 $\mu m$]{%
    \includegraphics[width=0.32\textwidth]{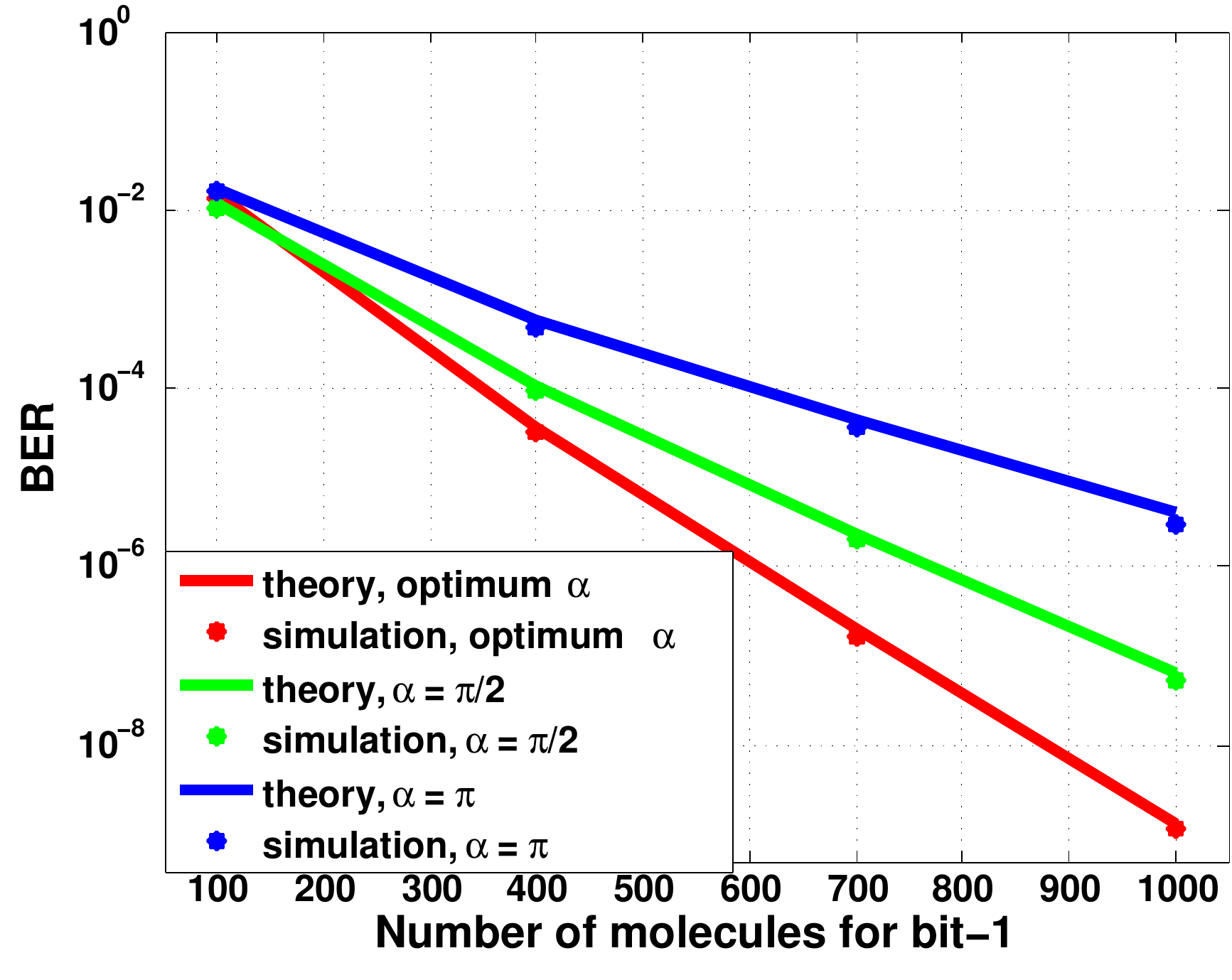}}
    \label{fig:aa}\hfill
    \subfloat[BER curves for $t_s$=300ms and $d$=7 $\mu m$]{%
    \includegraphics[width=0.32\textwidth]{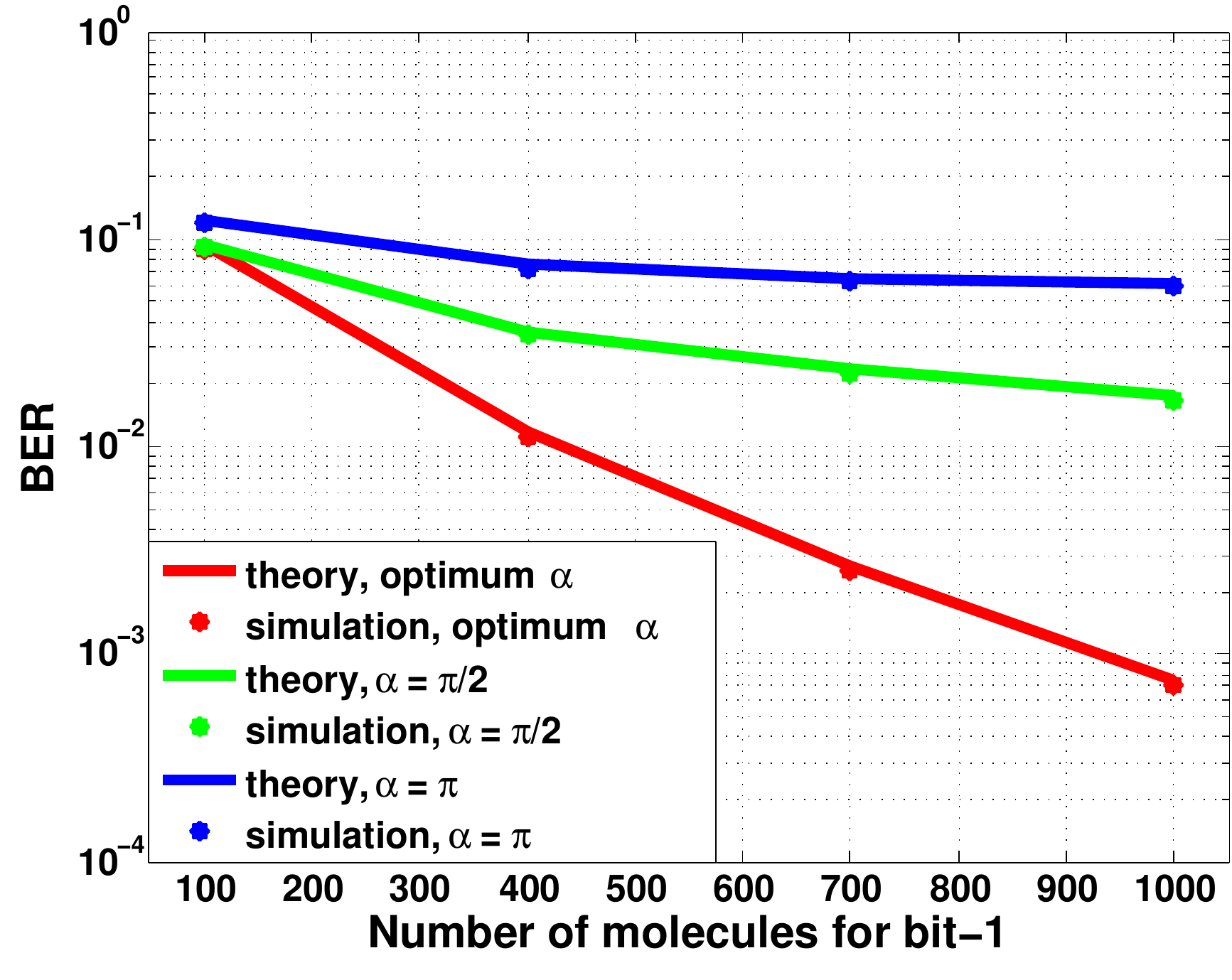}}
    \label{fig:bbt}\hfill 
    \subfloat[BER curves for $t_s$=400ms and $d$=7 $\mu m$]{%
    \includegraphics[width=0.32\textwidth]{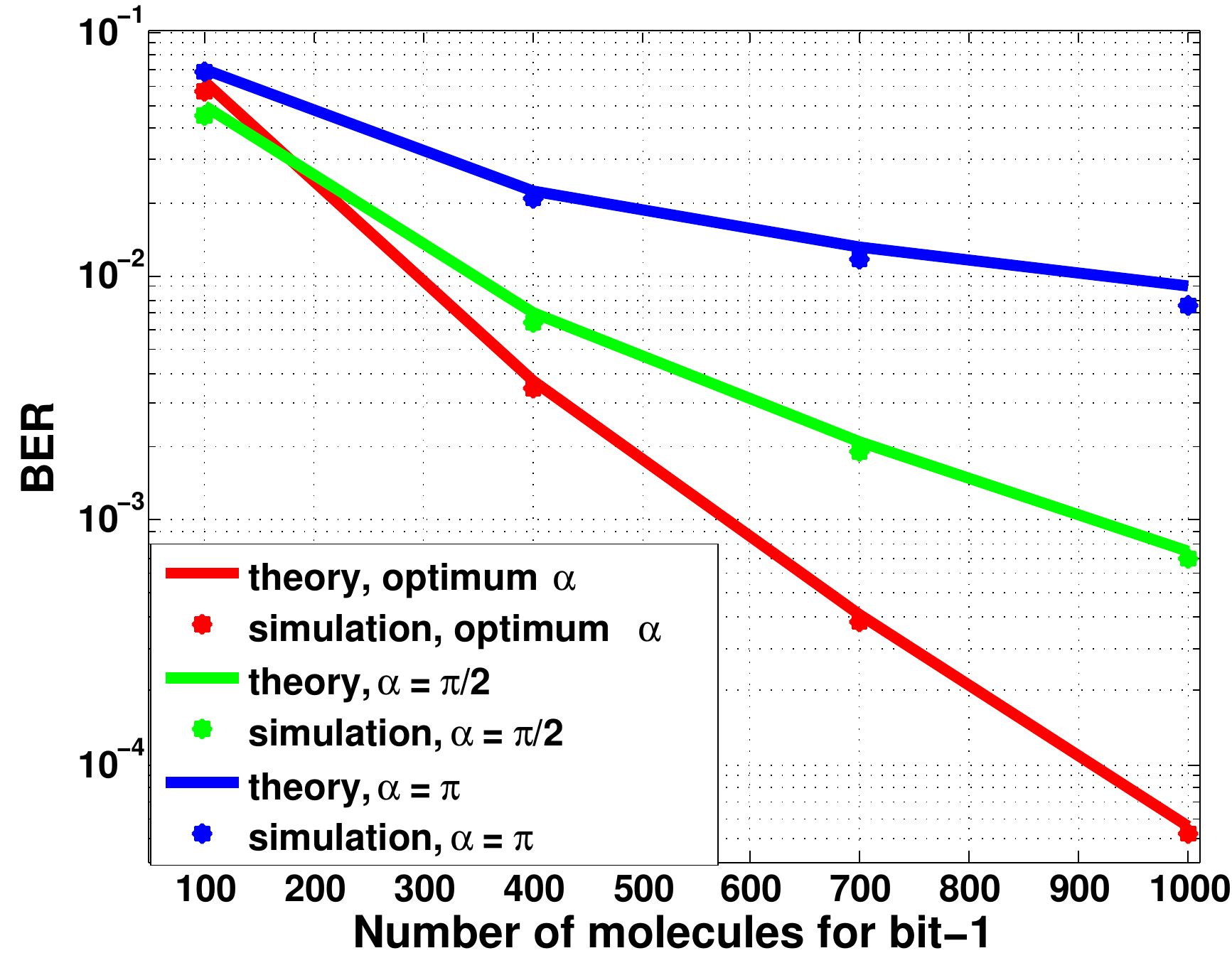}}
    \label{fig:bbtx}\\ 
    \subfloat[channel taps for $t_s$=300ms and $d$=5 $\mu m$]{%
    \includegraphics[width=0.32\textwidth]{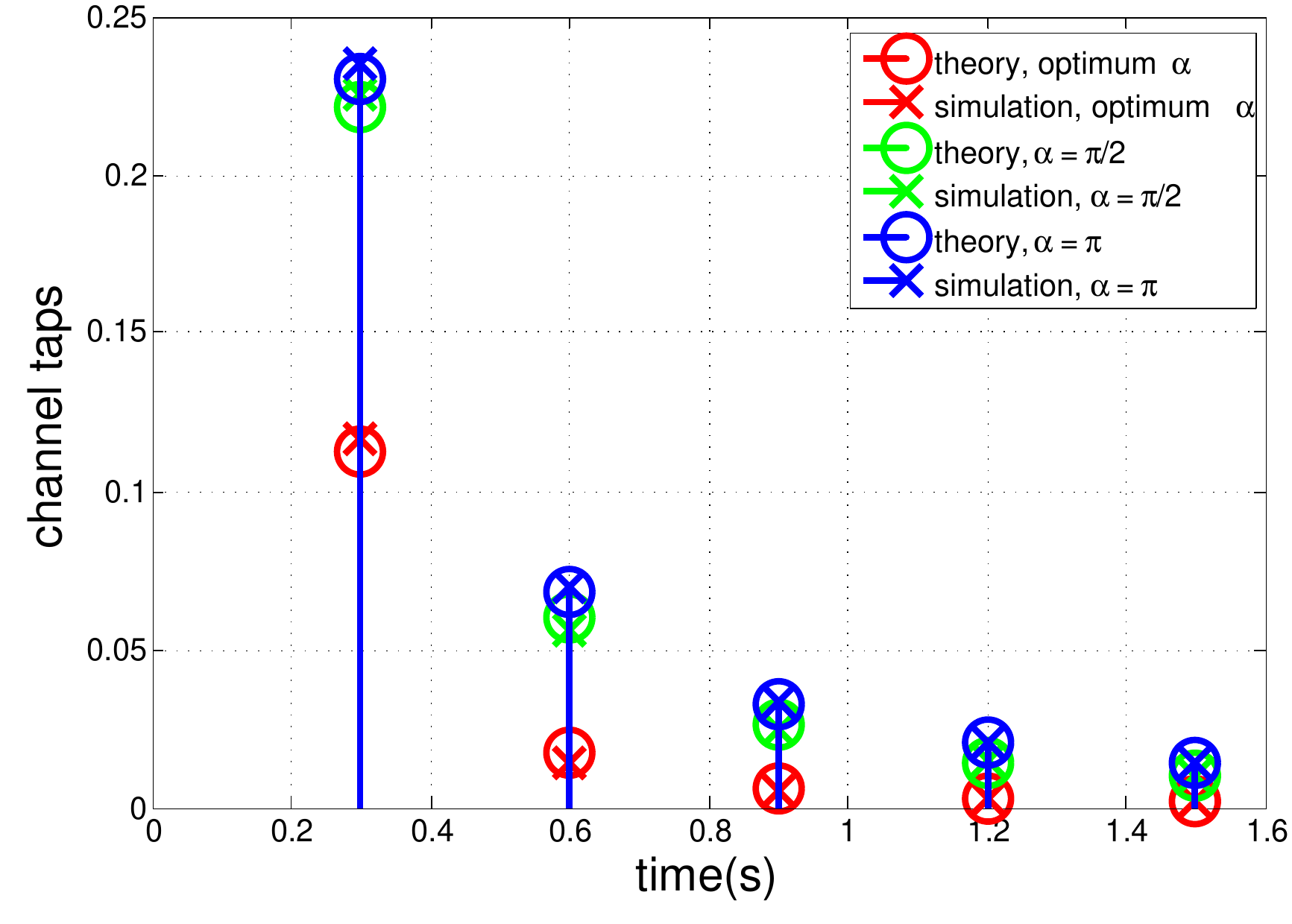}}
    \label{fig:aaa}\hfill
    \subfloat[channel taps for $t_s$=300ms and $d$=7 $\mu m$]{%
    \includegraphics[width=0.32\textwidth]{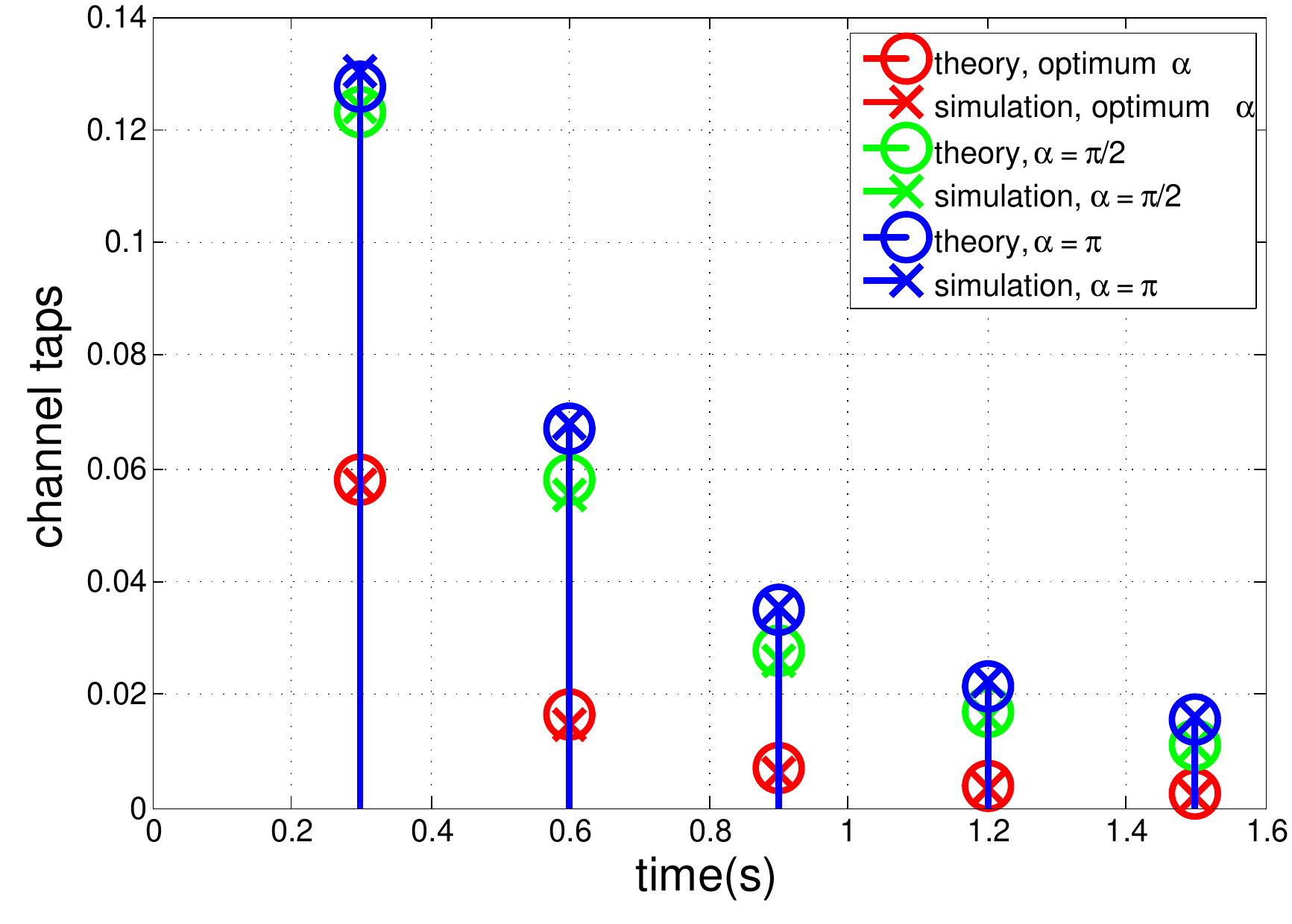}}
    \label{fig:bb}\hfill 
    \subfloat[channel taps for $t_s$=400ms and $d$=7 $\mu m$]{%
    \includegraphics[width=0.32\textwidth]{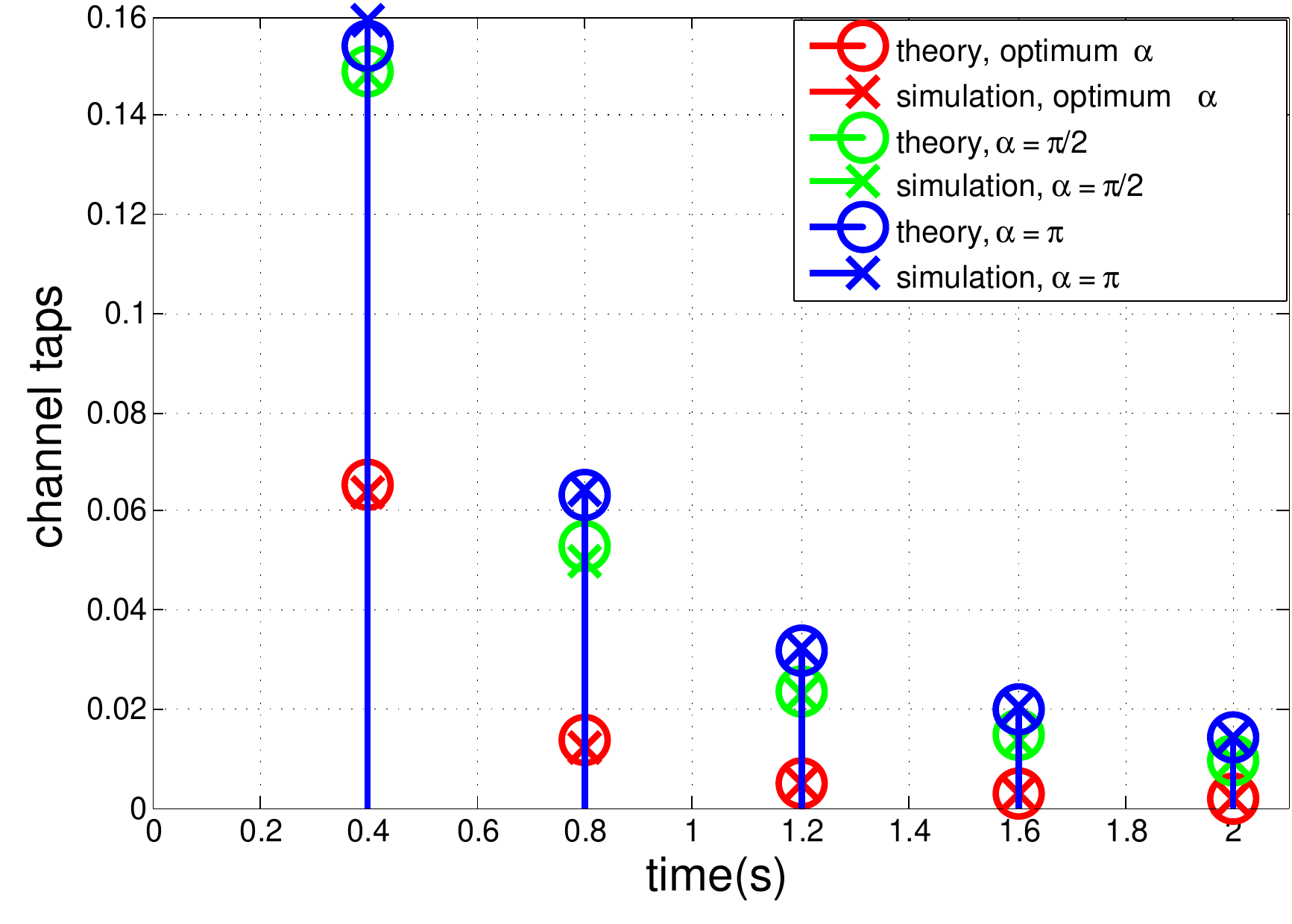}}
    \label{fig:bbx}
\caption{Top: BER vs number of molecules per bit-1 ($M$) curves  for $r_r$=5 $\mu m$,  $D=80 { \mu { m }^{ 2 } }/{ s }$ with different $d$=$r_0$-$r_r$  and $t_s$ values with receiver  $\alpha$=180$^{\circ}$ (convetional CSK), $\alpha$=90$^{\circ}$ and optimum $\alpha$ to obtain lowest BER. Bottom: corresponding channel taps of the communication systems }
\label{fig:berANDtaps}
\end{figure*}

 {In Fig.~\ref{fig:berANDtaps}, we present} the BER curves of three systems; $\alpha=\pi$ (conventional receiver), $\alpha=\pi/2$ (half sphere), and $\alpha=\optangle$ as well as their corresponding channel taps \tbirkan{for both simulation and analytical results}. Considering this figure, it can be concluded that the performance of the system will be significantly improved if  $\alpha$ is chosen properly. Although the signal tap is also decreased with this method, due to the decrease in ISI, this reduction is compensated. Furthermore, it can also be seen that analytical solutions using \eqref{eq:floatingeq2} and simulations are coherent. 

 



\section{Conclusion\label{sec_conclusion}}

{ In this paper, it has been confirmed that a partially counting absorbing receiver demonstrates a significant improvement over the conventional fully absorbing one. Due to the nature of diffusion, it can be expected that the molecules received in the back lobe of the receiver will most possibly take longer time to reach that point than the molecules received in the front lobe. Therefore, the molecules absorbed in the back lobe most likely belong to the previous transmitted symbols. Thus, they contribute to ISI. We, therefore, have proposed a counting region on the spherical receiver surface that faces towards the transmitter node. In order to justify this idea, we have derived the joint cumulative angle and the time distribution of the absorbed molecules at the receiver surface that had yet to be derived in the literature. Using this function and simulations, we have observed that the molecules are likely to be accumulated with a certain range of angles, which satisfies our claim. We, then, have examined the received signal model for various parameters. The optimum counting region to obtain the lowest BER was also derived. We have presented here evidence of the improved performance of the proposed system. As future work, our plans are to weight counting regions by an optimization approach so as to improve the performance of the system even better than how it did here. We intend to adopt this work to nanonetworks that involve one hub and many transmitters that aim to send their messages to this hub, the counting region of which should be assigned to the transmitters using the concepts proposed in this work.}

\section*{APPENDIX}
\label{app}

\tbayram{The objective function $SID=\left[{ 2F\left( \alpha ,t_s \right) -F\left( \alpha ,\infty \right) }  \right]$ can be written explicitly using \eqref{eq:floatingeq2} and discarding the $\alpha$ independent terms as
}
\begin{align}
 SID=&\frac { Y\left ( a {\erfc}\left( \frac { \sqrt { x }  }{ \sqrt { 4a }  }  \right) +\frac { 1 }{ \sqrt { 2\pi  }  } \sqrt { a } \sqrt { x } { Ei }\left( -\frac { \left( x \right)  }{ 4a }  \right)  \right)  }{ a\sqrt { x }  } +\frac { M }{ \sqrt { x }  } \\
&=SID_1 +\frac { M }{ \sqrt { x }  },
\end{align}
\tbayram{where $x=r_{ 0 }^{ 2 }-2r_{ 0 }{ r }_{ r }\cos  (\alpha )+{ r }_{ r }^2$, $a=D t_s$, $Y=\frac{-2 r_r \fhitx{r_0}{r_r}}{U(t_s)}$, and $M={ { r }_{ 0 }^{ 2 } }/{ 2 }$. Before taking the derivative of $SID$ with respect to $\alpha$, it needs to be simplified. $SID_1$ can be expanded to the series around $x=0$, and, since $x$ is on the order $10^{-12}$, higher order terms can be neglected. Therefore,  $SID_1$ can be written as  $SID_{ 1 }\approx \frac { Y }{ \sqrt { x }  } -\frac { Y\log { x }  }{ 2\sqrt { \pi a }  }$. Using this approximation, $SID$ can be written as} 

\begin{align}
 SID \approx \frac { Y }{ \sqrt { x }  } -\frac { Y\log { x }  }{ 2\sqrt { \pi a }  } +\frac { M }{ \sqrt { x }  }.
\end{align}

Taking the derivative of $SID$ with respect to $\alpha$ and equating it to 0, we can arrive at

\begin{align}
\frac { Y }{ \sqrt { \pi a } (r_{ 0 }^{ 2 }-2r_{ 0 }{ r }_{ r }\cos  (\alpha )+{ r }_{ r }^2) } -\frac { M+Y }{ { (r_{ 0 }^{ 2 }-2r_{ 0 }{ r }_{ r }\cos  (\alpha )+{ r }_{ r }^2) }^{ 1.5 } } =0.
\label{optim}
\end{align}

Hence, $\optangle$ can be obtained by solving \eqref{optim} as

\begin{align}
\optangle=\cos ^{ -1 }{ \left( \frac { { r }_{ 0 }^{ 2 }+{ r }_{ r }^{ 2 }-{ \left( \cfrac { \sqrt { \pi a } (Y+M) }{ Y }  \right)  }^{ 2 }\quad  }{ 2{ r }_{ 0 }{ r }_{ r } }  \right)  }.
\end{align}

\section*{ACKNOWLEDGEMENT}
The work of and H.B. Yilmaz, C.-B. Chae, T. Tugcu, and A.E. Pusane was supported in part by the joint project titled MEDUSA between TUBITAK of Turkey and NRF of South Korea. The work of T. Tugcu was also partially supported by the State Planning Organization (DPT) of Turkey under the project TAM (2007K120610).

\bibliographystyle{IEEEtran}
\bibliography{IEEEabrv,cap_rx_bib_file}

\end{document}